\documentclass[10pt,twocolumn,english,pra,showpacs,eqsecnum,floatfix]{revtex4}
\usepackage[T1]{fontenc}
\usepackage[latin9]{inputenc}
\usepackage{bm}
\usepackage{amsmath}
\usepackage{graphicx}
\usepackage{amssymb}

\usepackage{graphicx}
\usepackage{amssymb}
\usepackage{times}
\usepackage{amsmath}
\usepackage{amsthm}

\def\x#1{\chi^{(#1)}}

\usepackage{babel}

\begin{document}

\title{Resonant Cascaded Down-Conversion}

\author{Christian Weedbrook, Ben Perrett and Karen V. Kheruntsyan}

\affiliation{ARC Centre of Excellence for Quantum-Atom Optics, Department of Physics,
University of Queensland, Brisbane, Qld 4072, Australia}

\author{Peter D. Drummond}

\affiliation{ARC Centre of Excellence for Quantum-Atom Optics, Centre for Atom
Optics and Ultrafast Spectroscopy, Swinburne University of Technology,
Melbourne, Vic 3122, Australia}

\author{Raphael C. Pooser and Olivier Pfister}

\affiliation{Department of Physics, University of Virginia, 382 McCormick Road,
Charlottesville, VA 22904-4714, USA}

\date{\today}
\begin{abstract}
We analyze an optical parametric oscillator (OPO) in which cascaded
down-conversion occurs inside a cavity resonant for all modes but
the initial pump. Due to the resonant cascade design, the OPO present
two $\x2$ level oscillation thresholds that are therefore remarkably
lower than for a $\x3$ OPO. This is promising for reaching the regime
of an effective third-order nonlinearity well above both thresholds.
Such a $\x2$ cascaded device also has potential applications in frequency
conversion to far infra-red regimes. But, most importantly, it can
generate novel multi-partite quantum correlations in the output radiation,
which represent a step beyond squeezed or entangled light. The output
can be highly non-Gaussian, and therefore not describable by any semi-classical
model.
\end{abstract}

\pacs{03.67.-a, 42.50.Lc, 03.65.Ta, 03.67.Mn, 42.65.Yj}

\maketitle

\section{Introduction}

Continuous-variable (CV) quantum information is an interesting flavor
of quantum information (QI) \cite{book,rmp}. While easily implemented
by use of well established quantum optics techniques, benefiting from
large flow rates and broad spectral band, it has long been based on
coherent states and linear Bogoliubov transformations (quadratic Hamiltonians)
and therefore restricted to positive Wigner functions of Gaussian
character. These states are not general enough for universal quantum
information operations\cite{llb}. For instance, it has been shown
that quantum computation based solely on Gaussian CV states can be
efficiently simulated by a classical computer \cite{eff_cl_sim}.
Also, CV entanglement purification requires a Kerr-nonlinearity-based
QND measurement \cite{duan} or, in general, a non-Gaussian state
\cite{plenio}. However, it has also been shown that one-way quantum
computing can be implemented using Gaussian cluster-state entanglement
combined with non-CV (e.g., photon counting) measurements \cite{menicucci}.

Recently, successful {}``degaussification\textquotedbl{} experiments,
using homodyne detection conditioned on single-photon detection \cite{lvovsky,wenger,bellini,polzik},
have successfully generated negative Wigner functions from initial
squeezed states. Here, we investigate different type of sources, which
can produce non-Gaussian light directly. Theoretical studies of optical
parametric oscillators (OPO), which are based on a single second-order
optical nonlinearity ($\x2$) have shown non-Gaussian signatures to
be rather scarce \cite{drummond_deg} except in the case of the tripartite
correlations between the three fields \cite{drummond_nondeg}. An
interesting approach is to use an optical nonlinearity of, at least,
third order. This has been theoretically investigated \cite{felbinger,braunstein,hillery,banaszek}.
In practice, a $\x3$ based OPO would have the problems of requiring
a very large and possibly prohibitive input power threshold for downconversion,
together with an even higher threshold for the onset of nonclassical
effects, such as the formation of star states \cite{felbinger}.

In this paper, we show how the use of a {\em cavity-resonant} cascade
of second-order nonlinearities can yield a low-threshold OPO which
possesses the effective behavior of a $\x3$ OPO in certain regimes
and is more accessible experimentally. Note that related systems have
been studied before, in the purely classical case and for completely
different purposes, such as producing new tunable optical sources
in the infrared \cite{moore} or achieving optical phase-locking in
a 3:1 frequency ratio for frequency metrology \cite{Zondy_PRA2001,Zondy_PRL2004}.
Parametric amplifiers and oscillators have become a widely used, even
standard part of the repertoire of laser physics and quantum optics
\cite{Yar89}. Above the classical threshold points, these devices
are a useful tool for frequency conversion. Below threshold, quantum
effects dominate, leading to squeezing and entanglement. These devices
that rely on non-resonant, nonlinear optics interactions have proved
experimentally superior to other resonant or near-resonant alternatives,
due to the fact that absorption is suppressed.

There are other possible quantum effects available, as well as direct
down-conversion in the linear regime well below threshold. For example,
exploration of non-equilibrium quantum criticality is possible near
threshold. This results in large critical fluctuations and phase-transitions.
The fluctuations in this case become non-Gaussian, but the dominant
critical fluctuations have a rather classical character. Here, we
explore another path to such non-Gaussian behavior, in which extremely
nonclassical correlations are generated through the presence of a
second down-conversion crystal placed inside the cavity. We show that
this results in an intricate pattern of new phase-transitions at the
classical level, in which there are two distinct threshold points.
At the quantum level, below the first threshold, there are very strong
triple correlations between the three down-converted modes, which
have no classical analog.

This paper is structured as follows. In Section II, we explain the
basic model and the theoretical phase-space techniques that are used
here. In Section III, we present an analytical study of the system's
stationary solutions. In Section IV, we turn to a treatment of stability
properties and fluctuations in one particular type of down-conversion
scenario. In Section V, we give numerical simulations of more general
cases which can also yield regimes of interest. These simulations
demonstrate the stability regions, in the same spirit as was achieved
for the $\x2$ OPO \cite{lugiato}. We give conclusions in Section
VI.


%
\begin{figure}[!ht]
\begin{centering}
\includegraphics[width=8cm]{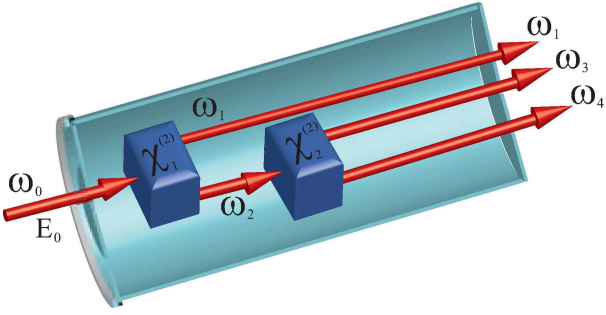} 
\par\end{centering}

\caption{(Color online) Schematic of the resonant cascaded down-conversion
system. A driving field, that is pumped at a frequency $\omega_{0}$
with amplitude $E_{0}$, enters a cavity that contains two $\chi^{(2)}$
nonlinear crystals. The first $\chi_{1}^{(2)}$ crystal down-converts
the original mode $\hat{a}_{0}$ into two modes $\hat{a}_{1}$ and
$\hat{a}_{2}$ with frequencies $\omega_{1}$ and $\omega_{2}$, respectively.
Then the second mode $\hat{a}_{2}$ undergoes a further down-conversion,
via the $\chi_{2}^{(2)}$ crystal, into the two modes $\hat{a}_{3}$
and $\hat{a}_{4}$ with frequencies $\omega_{3}$ and $\omega_{4}$,
respectively.}

\centering{}\label{schematic} 
\end{figure}

\section{Analytical treatment}

\subsection{Cascaded parametric oscillator model}

The model system for cascaded down-conversion consists of two quadratically
nonlinear elements with nonlinearities $\chi_{1}$ and $\chi_{2}$
inside an optical cavity (c.f. Fig.~\ref{schematic}). The cavity
supports five resonant modes at frequencies $\omega_{i}$ ($i=0,1,...4$).
The mode $\omega_{0}$ is the pump mode, driven by an external coherent
driving field at the same frequency $\omega_{0}$. The cavity modes
$\omega_{i}$ are described by creation and annihilation operators
$\hat{a}_{i}^{\dagger}$ and $\hat{a}_{i}$ with commutation relations
$[\hat{a}_{i},\hat{a}_{j}^{\dagger}]=\delta_{ij}$. The first nonlinear
element converts the pump mode $\omega_{0}$ into the signal and idler
modes $\omega_{1}$ and $\omega_{2}$ by means of nondegenerate parametric
down-conversion, where $\omega_{0}=\omega_{1}+\omega_{2}$ ($\omega_{1}\neq\omega_{2}$).
The second nonlinear crystal supports down-conversion of the mode
$\omega_{2}$ into the second pair of signal and idler modes, $\omega_{3}$
and $\omega_{4}$, where $\omega_{2}=\omega_{3}+\omega_{4}$. We will
call the field $\hat{a}_{2}$ at $\omega_{2}$ the {}``intermediate
pump.\textquotedbl{} The modes may decay via cavity losses at the
respective rates $\gamma_{i}$, $i\in[0,4]$.

In the absence of the optical cavity, this interaction constitutes
a cascade of quantum systems in the sense investigated by several
authors before \cite{gardiner}, where the second stage does not feed
back to the first stage. Here, the situation is different precisely
because of the cavity feedback, hence our use of the term {\em resonant
cascade} throughout the paper. Within this frame, we will distinguish
two situations: the first one is the {\em nondegenerate} resonant
cascade, for which the fields $\hat{a}_{1}$, $\hat{a}_{3}$, and
$\hat{a}_{4}$ are distinguishable (i.e., \ $\omega_{1}\neq\omega_{3}\neq\omega_{4}$,
or having different polarizations or wave-vector directions). In this
case, the only physical observable affected by both stages of the
cascade is the intermediate pump $\hat{a}_{2}$. This is the case
that will be investigated analytically, with additional simplifying
hypotheses, and numerically, without those hypotheses. The second
case is the {\em degenerate} resonant cascade, for which the signal
fields are indistinguishable: $\hat{a}_{1}\equiv\hat{a}_{3}\equiv\hat{a}_{4}$
(and hence $\omega_{1}=\omega_{3}=\omega_{4}$). In that case the
signal field and the intermediate pump interact in both nonlinear
media and the dynamics are richer. That case will be explored by numerical
simulations. Obviously, intermediate situations do also exist, e.g.,
$\omega_{1}=\omega_{3}\neq\omega_{4}$, but we will not consider them
here.

\subsection{Hamiltonian and equations of motion}

The model Heisenberg-picture Hamiltonian for the system, in the rotating-wave
approximation, is given by: \begin{align}
\hat{H}= & \sum_{i=0}^{4}\hbar\omega_{i}\hat{a}_{i}^{\dagger}\hat{a}_{i}+i\hbar(E_{0}e^{-i\omega_{0}t}\hat{a}_{0}^{\dagger}-E_{0}^{\ast}e^{i\omega_{0}t}\hat{a}_{0})\nonumber \\
 & +i\hbar\chi_{1}(\hat{a}_{0}\hat{a}_{1}^{\dagger}\hat{a}_{2}^{\dagger}-\hat{a}_{0}^{\dagger}\hat{a}_{1}\hat{a}_{2})+i\hbar\chi_{2}(\hat{a}_{2}\hat{a}_{3}^{\dagger}\hat{a}_{4}^{\dagger}-\hat{a}_{2}^{\dagger}\hat{a}_{3}\hat{a}_{4})\nonumber \\
 & +\sum_{i=0}^{4}(\hat{a}_{i}\hat{\Gamma}_{i}^{\dagger}+\hat{a}_{i}^{\dagger}\hat{\Gamma}_{i})\label{Hamiltonian}\end{align}
 Here, $E_{0}$ describes the complex amplitude of the driving field.
The coupling constants $\chi_{1}$ and $\chi_{2}$ are proportional
to the second-order susceptibilities of the two nonlinear elements,
respectively. We assume that they are positive, without loss of generality,
since phase factors can always be absorbed into the definitions of
the mode functions and their operators. The operators $\hat{\Gamma}_{i}$
and $\hat{\Gamma}_{i}^{\dagger}$ describe the coupling of each intracavity
mode to the reservoir of external modes. These give rise to the losses
of the cavity modes $\omega_{i}$ at rates $\gamma_{i}$.

\subsubsection{Master Equation}

Transforming to an interaction picture in which all operators are
transformed to rotating frames, i.e., \begin{equation}
\hat{a}_{j}(t)=\hat{a}_{j}e^{-i\omega_{j}t},\end{equation}
 one can derive the following master equation for the system density
operator $\hat{\rho}$ \cite{Louisell}: \begin{align}
\frac{\partial\hat{\rho}}{\partial t} & =[E_{0}\hat{a}_{0}^{\dagger}-E_{0}^{\ast}\hat{a}_{0},\hat{\rho}]+\chi_{1}[\hat{a}_{0}\hat{a}_{1}^{\dagger}\hat{a}_{2}^{\dagger}-\hat{a}_{0}^{\dagger}\hat{a}_{1}\hat{a}_{2},\hat{\rho}]\nonumber \\
 & +\chi_{2}[\hat{a}_{2}\hat{a}_{3}^{\dagger}\hat{a}_{4}^{\dagger}-\hat{a}_{2}^{\dagger}\hat{a}_{3}\hat{a}_{4},\hat{\rho}]\nonumber \\
 & +\sum_{i=0}^{4}\gamma_{i}(2\hat{a}_{i}\hat{\rho}\hat{a}_{i}^{\dagger}-\hat{\rho}\hat{a}_{i}^{\dagger}\hat{a}_{i}-\hat{a}_{i}^{\dagger}\hat{a}_{i}\hat{\rho}).\label{master equation}\end{align}
 While in principle this master equation can be solved numerically
in a number-state representation, in practice this is not possible.
The complexity of the Hilbert space --- especially for this five mode
problem --- is enormous, given any moderate number of photons present
in the five interacting modes. Instead, we solve this problem using
phase-space representation methods, such as the positive-P representation\cite{gen-p-rep}.

\subsection{Positive-P representation}

Using the positive-P representation we can transform the master equation,
Eq.~(\ref{master equation}), into a Fokker-Planck equation \cite{gen-p-rep}
expressed as: \begin{align}
\frac{\partial}{\partial t}P(\mathbf{\alpha},\mathbf{\alpha}^{+},t)= & \left[\frac{\partial}{\partial\alpha_{0}}(\gamma_{0}\alpha_{0}-E_{0}+\chi_{1}\alpha_{1}\alpha_{2})\right.\nonumber \\
 & +\frac{\partial}{\partial\alpha_{1}}(\gamma_{1}\alpha_{1}-\chi_{1}\alpha_{0}\alpha_{2}^{+})\nonumber \\
 & +\frac{\partial}{\partial\alpha_{2}}(\gamma_{2}\alpha_{2}-\chi_{1}\alpha_{0}\alpha_{1}^{+}+\chi_{2}\alpha_{3}\alpha_{4})\nonumber \\
 & +\frac{\partial}{\partial\alpha_{3}}(\gamma_{3}\alpha_{3}-\chi_{2}\alpha_{2}\alpha_{4}^{+})\nonumber \\
 & +\frac{\partial}{\partial\alpha_{4}}(\gamma_{4}\alpha_{4}-\chi_{2}\alpha_{2}\alpha_{3}^{+})\nonumber \\
 & +\frac{\partial^{2}}{\partial\alpha_{1}\partial\alpha_{2}}(\chi_{1}\alpha_{0})\nonumber \\
 & \left.+\frac{\partial^{2}}{\partial\alpha_{3}\partial\alpha_{4}}(\chi_{2}\alpha_{2})+h.c.\right]P(\mathbf{\alpha},\mathbf{\alpha}^{+},t).\label{Fokker-Planck}\end{align}
 Here, $\mathbf{\alpha}\equiv(\alpha_{0},\alpha_{1},\alpha_{2},\alpha_{3},\alpha_{4})$
and $\mathbf{\alpha}^{+}\equiv(\alpha_{0}^{+},\alpha_{1}^{+},\alpha_{2}^{+},\alpha_{3}^{+},\alpha_{4}^{+})$
represent the sets of coherent state amplitudes $\alpha_{i}$ and
$\alpha_{i}^{+}$ in the expansion of the density operator in terms
of the positive $P$-representation, corresponding to the annihilation
and creation operators $\hat{a}_{i}$ and $\hat{a}_{i}^{\dagger}$.
We recall that in the positive $P$-representation, the amplitudes
$\alpha_{i}$ and $\alpha_{i}^{+}$ are independent complex $c$-numbers,
and h.c. in Eq.~(\ref{Fokker-Planck}) represents the terms equivalent
to Hermitian conjugate operators, obtained from the previous terms
by replacing $\alpha_{j}\rightarrow\alpha_{j}^{+}$ and vice versa,
while $E_{0}$ is replaced by $E_{0}^{\ast}$. The transformation
requires an assumption of vanishing boundary terms which can be checked
numerically. This is generally extremely well-satisfied\cite{GGD}
for these open systems provided $\chi_{i}<<\gamma_{j}$, which is
typically the case in nonlinear optics experiments. If required, further
stochastic gauge transformations\cite{Deuar} can be used to eliminate
boundary terms.

The Fokker-Planck equation (\ref{Fokker-Planck}) is equivalent to
the following set of stochastic differential equations \cite{Wal08},
in the It${\rm \hat{o}}$ form: \begin{align}
\dot{\alpha}_{0} & =-\gamma_{0}\alpha_{0}+E_{0}-\chi_{1}\alpha_{1}\alpha_{2},\nonumber \\
\dot{\alpha}_{1} & =-\gamma_{1}\alpha_{1}+\chi_{1}\alpha_{0}\alpha_{2}^{+}+\sqrt{\chi_{1}\alpha_{0}}\zeta_{1}(t),\nonumber \\
\dot{\alpha}_{2} & =-\gamma_{2}\alpha_{2}+\chi_{1}\alpha_{0}\alpha_{1}^{+}-\chi_{2}\alpha_{3}\alpha_{4}+\sqrt{\chi_{1}\alpha_{0}}\zeta_{2}(t),\nonumber \\
\dot{\alpha}_{3} & =-\gamma_{3}\alpha_{3}+\chi_{2}\alpha_{2}\alpha_{4}^{+}+\sqrt{\chi_{2}\alpha_{2}}\zeta_{3}(t),\nonumber \\
\dot{\alpha}_{4} & =-\gamma_{4}\alpha_{4}+\chi_{2}\alpha_{2}\alpha_{3}^{+}+\sqrt{\chi_{2}\alpha_{2}}\zeta_{4}(t).\label{eq of motion}\end{align}
 together with the corresponding equations for $\dot{\alpha}_{i}^{+}$.
Here, the dots imply a time derivative, and the terms $\zeta_{i}(t)$
and $\zeta_{i}^{+}(t)$ are independent complex Gaussian noise sources
with zero means and the following nonzero correlations: \begin{align}
\left\langle \zeta_{1}(t)\zeta_{2}(t^{\prime})\right\rangle  & =\left\langle \zeta_{1}^{+}(t)\zeta_{2}^{+}(t^{\prime})\right\rangle =\delta(t-t^{\prime}),\nonumber \\
\left\langle \zeta_{3}(t)\zeta_{4}(t^{\prime})\right\rangle  & =\left\langle \zeta_{3}^{+}(t)\zeta_{4}^{+}(t^{\prime})\right\rangle =\delta(t-t^{\prime}).\label{correlations}\end{align}
 The above set of the stochastic equations of motion, Eq.~(\ref{eq of motion}),
can be solved either numerically or else using approximate analytic
treatments such as perturbation expansions around stable semi-classical
steady states. Quantum mechanical observables that are expressed in
terms of normally ordered operator moments $\left\langle (\hat{a}_{j}^{\dagger})^{n}(\hat{a}_{i})^{m}\right\rangle $
correspond to stochastic averages $\left\langle (\alpha_{i})^{m}(\alpha_{j}^{+})^{n}\right\rangle $.

\subsection{The semi-classical theory}

We can also transcribe the master equation, Eq.~(\ref{master equation}),
as a c-number phase space evolution equation using the Wigner representation
\cite{Leo} \begin{equation}
P_{W}(\bm{\alpha},\bm{\alpha}^{*})=\frac{1}{\pi^{2}}\int_{-\infty}^{\infty}d^{10}\bm{z}\;\chi_{W}(\bm{z},\bm{z}^{*})e^{-i\bm{z}^{*}\cdot\bm{\alpha}^{*}}e^{-i\bm{z}\cdot\bm{\alpha}}\label{a9}\end{equation}
 where $\chi_{S}(\bm{z},\bm{z}^{*})$, the characteristic function
for the Wigner representation , is given by \begin{equation}
\chi_{W}(\bm{z},\bm{z}^{*})=Tr\left(\rho e^{i\bm{z}^{*}\bm{a}^{\dagger}+i\bm{z}\cdot\bm{a}}\right)\label{a8}\end{equation}
 This transcription is particularly useful for semi-classical treatments
in which we include quantum noise terms from the reservoirs, but neglect
higher-order quantum noise from the nonlinear couplings. This approximation
is also called a truncated Wigner approximation, as it is obtained
from a full Wigner-Moyal equations via truncation of third-order derivatives. 

The equation for the Wigner function for the nondegenerate parametric
amplifier that corresponds to the master equation given by Eq.~(\ref{master equation})
turns out to be \cite{gardiner} \begin{eqnarray*}
\frac{\partial P_{W}}{\partial t} & = & \left\{ \frac{\partial}{\partial\alpha_{0}}\left(\gamma_{0}\alpha_{0}+\chi_{1}\alpha_{1}\alpha_{2}-{\mathcal{E}}\right)\right.\\
 &  & +\frac{\partial}{\partial\alpha_{0}^{*}}\left(\gamma_{0}\alpha_{0}^{*}+\chi_{1}\alpha_{1}^{*}\alpha_{2}^{*}-{\mathcal{E}}\right)\\
 &  & +\frac{\partial}{\partial\alpha_{1}}\left(\gamma_{1}\alpha_{1}-\chi_{1}\alpha_{2}^{*}\alpha_{0}\right)+\frac{\partial}{\partial\alpha_{1}^{*}}\left(\gamma_{1}\alpha_{1}^{*}-\chi_{1}\alpha_{2}\alpha_{0}^{*}\right)\\
 &  & +\frac{\partial}{\partial\alpha_{2}}\left(\gamma_{2}\alpha_{2}-\chi_{1}\alpha_{1}^{*}\alpha_{0}+\chi_{2}\alpha_{3}\alpha_{4}\right)\\
 &  & +\frac{\partial}{\partial\alpha_{2}^{*}}\left(\gamma_{2}\alpha_{2}^{*}-\chi_{1}\alpha_{1}\alpha_{0}^{*}+\chi_{2}\alpha_{3}^{*}\alpha_{4}^{*}\right)\\
 &  & +\frac{\partial}{\partial\alpha_{3}}\left(\gamma_{3}\alpha_{3}-\chi_{2}\alpha_{4}^{*}\alpha_{2}\right)+\frac{\partial}{\partial\alpha_{3}^{*}}\left(\gamma_{3}\alpha_{3}^{*}-\chi_{2}\alpha_{4}\alpha_{2}^{*}\right)\\
 &  & +\frac{\partial}{\partial\alpha_{4}}\left(\gamma_{4}\alpha_{4}-\chi_{2}\alpha_{3}^{*}\alpha_{2}\right)+\frac{\partial}{\partial\alpha_{4}^{*}}\left(\gamma_{4}\alpha_{4}^{*}-\chi_{2}\alpha_{3}\alpha_{2}^{*}\right)\\
 &  & +\gamma_{0}\frac{\partial^{2}}{\partial\alpha_{0}\partial\alpha_{0}^{*}}+\gamma_{1}\frac{\partial^{2}}{\partial\alpha_{1}\partial\alpha_{1}^{*}}+\gamma_{2}\frac{\partial^{2}}{\partial\alpha_{2}\partial\alpha_{2}^{*}}\\
 &  & +\gamma_{3}\frac{\partial^{2}}{\partial\alpha_{3}\partial\alpha_{3}^{*}}+\gamma_{4}\frac{\partial^{2}}{\partial\alpha_{4}\partial\alpha_{4}^{*}}\\
 &  & +\frac{\chi_{1}}{4}\left(\frac{\partial^{3}}{\partial\alpha_{1}\partial\alpha_{2}\partial\alpha_{0}^{*}}+\frac{\partial^{3}}{\partial\alpha_{1}^{*}\partial\alpha_{2}^{*}\partial\alpha_{0}}\right)\\
 &  & +\left.\frac{\chi_{2}}{4}\left(\frac{\partial^{3}}{\partial\alpha_{3}\partial\alpha_{4}\partial\alpha_{2}^{*}}+\frac{\partial^{3}}{\partial\alpha_{3}^{*}\partial\alpha_{4}^{*}\partial\alpha_{2}}\right)\right\} P_{W}\end{eqnarray*}
 It is common to drop the third order derivative terms, in an approximation
valid in the limit of large photon number. This allows one to equate
the resulting truncated, positive-definite Fokker-Planck equation
with a set of stochastic equations. These are: \begin{align}
\dot{\alpha}_{0} & =-\gamma_{0}\alpha_{0}+E_{0}-\chi_{1}\alpha_{1}\alpha_{2}+\sqrt{\gamma_{0}}\eta_{0}(t),\nonumber \\
\dot{\alpha}_{1} & =-\gamma_{1}\alpha_{1}+\chi_{1}\alpha_{0}\alpha_{2}^{*}+\sqrt{\gamma_{1}}\eta_{1}(t),\nonumber \\
\dot{\alpha}_{2} & =-\gamma_{2}\alpha_{2}+\chi_{1}\alpha_{0}\alpha_{1}^{*}-\chi_{2}\alpha_{3}\alpha_{4}+\sqrt{\gamma_{2}}\eta_{2}(t),\nonumber \\
\dot{\alpha}_{3} & =-\gamma_{3}\alpha_{3}+\chi_{2}\alpha_{2}\alpha_{4}^{*}+\sqrt{\gamma_{3}}\eta_{3}(t),\nonumber \\
\dot{\alpha}_{4} & =-\gamma_{4}\alpha_{4}+\chi_{2}\alpha_{2}\alpha_{3}^{*}+\sqrt{\gamma_{4}}\eta_{4}(t).\label{Wignerstochastic}\end{align}
 together with the corresponding equations for $\dot{\alpha}_{i}^{+}$.
Here, the conjugate equations have conjugate noises as in a normal
classical phase-space. The terms $\eta_{i}(t)$ are complex Gaussian
noise sources with zero means and the following nonzero correlations:
\[
\left\langle \eta_{i}(t)\eta_{j}^{*}(t^{\prime})\right\rangle =\delta_{ij}\delta(t-t^{\prime}),\]
 If we compare the two sets of It${\rm \hat{o}}$ stochastic equations,
we see that the noise terms in the positive-P equations, Eq.~(\ref{eq of motion}),
depend on the nonlinear coupling constant, while those in the Wigner
representation, Eq.~(\ref{Wignerstochastic}), do not.

The truncated Wigner theory can be regarded as a kind of hidden-variable
theory, since it behaves as though the non-commuting quadrature variables
were simple classical objects. These equations imply that $\langle\alpha_{i}\alpha_{i}^{\dagger}\rangle=\langle\hat{n}_{i}\rangle=1/2$
when there is no driving and no coupling, which is an expected result
in a symmetrically-ordered representation. However, the truncation
neglects third-order derivative terms which are present in the full
Wigner equation, and are not always negligible. The full Wigner theory
is equivalent to quantum mechanics, and has no such limitations but
it is no longer positive-definite, and therefore has no equivalent
stochastic formulation. The advantage of the positive-P method is
that it is able to generate stochastic equations without requiring
this questionable truncation approximation.

\section{Classical steady states}

We first analyze the classical steady states of the system and then
give the results of the linearized fluctuation analysis for their
stability in the next section. In the classical limit, all quantum
noise terms are neglected. The positive-P stochastic variables $\alpha_{i}$
and $\alpha_{i}^{+}$ are replaced by deterministic amplitudes $\alpha_{i}$
and $\alpha_{i}^{\ast}$, where $\alpha_{i}^{\ast}$ is the complex
conjugate of $\alpha_{i}$, and Eq.~(\ref{eq of motion}) then becomes
\begin{align}
\dot{\alpha}_{0} & =-\gamma_{0}\alpha_{0}+E_{0}-\chi_{1}\alpha_{1}\alpha_{2}\,,\nonumber \\
\dot{\alpha}_{1} & =-\gamma_{1}\alpha_{1}+\chi_{1}\alpha_{0}\alpha_{2}^{*},\nonumber \\
\dot{\alpha}_{2} & =-\gamma_{2}\alpha_{2}+\chi_{1}\alpha_{0}\alpha_{1}^{*}-\chi_{2}\alpha_{3}\alpha_{4},\nonumber \\
\dot{\alpha}_{3} & =-\gamma_{3}\alpha_{3}+\chi_{2}\alpha_{2}\alpha_{4}^{*},\nonumber \\
\dot{\alpha}_{4} & =-\gamma_{4}\alpha_{4}+\chi_{2}\alpha_{2}\alpha_{3}^{*}.\label{Classical}\end{align}
 The steady state solutions $\alpha_{i}^{0}$ are obtained from Eqs.
(\ref{Classical}) by putting all time derivatives equal to zero,
i.e., $\dot{\alpha}_{i}=0$. We only consider the steady-state solutions
in which $(\alpha_{i}^{0})^{+}=(\alpha_{i}^{0})^{\ast}$, as these
correspond to classical fields. This corresponds to neglecting the
effects of quantum fluctuations and considering the equations for
the mean field amplitudes $\alpha_{i}=\left\langle \hat{a}_{i}\right\rangle $,
assuming that higher-order correlations factorize.

The stability of the classical steady states with respect to small
fluctuations can be checked by deriving the linearized equations of
motion for the fluctuations $\delta\alpha_{i}(t)=\alpha_{i}(t)-\alpha_{i}^{0}$
and $\delta\alpha_{i}^{+}(t)=\alpha_{i}^{+}(t)-(\alpha_{i}^{0})^{\ast}$.
The steady states are stable provided all the eigenvalues of the appropriate
drift matrix of the linearized equations have negative real parts.
Here, we assume the following matrix form of the deterministic part
of the linearized equations of motion:

\begin{equation}
\mathbf{\dot{x}}=\mathbf{Ax}\,\,,\end{equation}
 where $\mathbf{A}$ is the drift matrix, and $\mathbf{x}$ denotes
a column vector for fluctuations $\{\delta\alpha_{i},\delta\alpha_{i}^{+}\}$.
If the linearized eigenvalue analysis reveals eigenvalues with non-negative
real parts, this implies that the steady states are unstable. In this
case, the linearized treatment of fluctuations around the classical
steady states cannot be employed, and the equations of motion have
to be treated exactly.

To simplify our analysis and make analytic solutions available, we
will assume that the damping rates for all modes except the pump mode
are equal to each other, \begin{equation}
\gamma_{1}=\gamma_{2}=\gamma_{3}=\gamma_{4}\equiv\gamma,\label{gammapprox}\end{equation}
 while the pump mode is strongly damped, 

\begin{equation}
\gamma_{0}>>\gamma,\label{gammazero}\end{equation}
to model an interferometer which is not resonant at the pump wavelength.
For simplicity we suppose that the coupling constants $\chi_{1}$
and $\chi_{2}$ are also equal: \begin{equation}
\chi_{1}=\chi_{2}\equiv\chi.\end{equation}
 Analysis of the equations of motion for the classical steady states
reveals three different types of solutions, corresponding to three
regimes of operation.

\begin{figure}
\includegraphics[width=8cm]{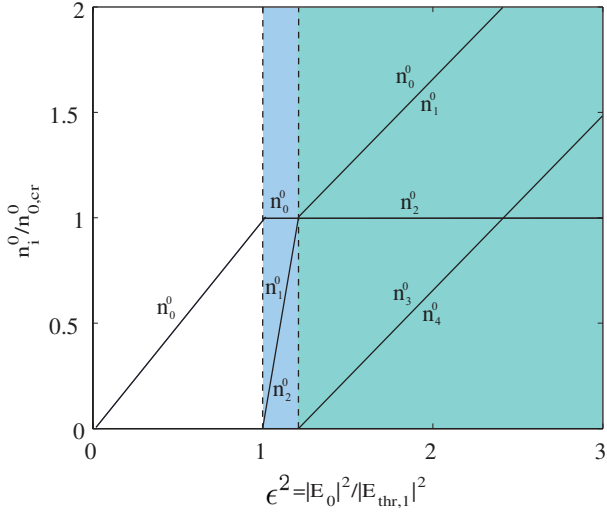} 

\caption{(Color online) Steady state solutions for the scaled intensities $n_{i}^{0}/n_{0,cr}^{0}$
of different modes ($i=0,1,...,4$) as a function of the driving field
intensity parameter $\epsilon^{2}\equiv\left|E_{0}\right|^{2}/\left|E_{thr,1}\right|^{2}$,
for $\gamma_{0}/\gamma=10$. Here, $n_{0,cr}^{0}=\gamma^{2}/\chi^{2}$
is the critical value of $n_{0}^{0}$ at the first threshold, $\left|E_{0}\right|^{2}=\left|E_{thr,1}\right|^{2}$
$\left(\epsilon_{thr,1}^{2}=1\right)$. The second threshold here
corresponds to $\epsilon_{thr,2}^{2}=(1+\gamma/\gamma_{0})^{2}=1.21$. }

\label{Fig: sss} 
\end{figure}

\subsubsection{Below threshold regime}

Here, the amplitudes of all intracavity modes except the pump mode
$\omega_{0}$ are zero, and we find that \begin{align}
 & \alpha_{1}^{0}=\alpha_{2}^{0}=\alpha_{3}^{0}=\alpha_{4}^{0}=0,\nonumber \\
 & \alpha_{0}^{0}=\frac{E_{0}}{\gamma_{0}}.\label{SS1}\end{align}
 The last equation can be rewritten in terms of the steady state intensity
$n_{0}^{0}=$ $\left|\alpha_{0}^{0}\right|^{2}$ (in photon number
units) and phase $\phi_{0}^{0}$ (where $\alpha_{0}^{0}=\sqrt{n_{0}^{0}}\exp(i\phi_{0}^{0})$):
\begin{align}
n_{0}^{0} & =\left|E_{0}\right|^{2}/\gamma_{0}^{2},\;\nonumber \\
\phi_{0}^{0} & =\varphi_{0},\end{align}
 where $\varphi_{0}$ is the phase of the driving field, i.e., $E_{0}=|E_{0}|\exp(i\varphi_{0})$.
The linearized stability analysis of these steady states (see Sec.
IV) reveals that they are stable for driving field intensities below
a certain critical (threshold) value, \begin{equation}
\left|E_{0}\right|^{2}<\left|E_{thr,1}\right|^{2},\label{region1}\end{equation}
 where \begin{equation}
\left|E_{thr,1}\right|^{2}\equiv\frac{\gamma_{0}^{2}\gamma^{2}}{\chi^{2}}\label{threshold1}\end{equation}
 is the first threshold. This allows us to introduce a dimensionless
relative driving field parameter, \[
\epsilon\equiv\frac{\left|E_{0}\right|}{\left|E_{thr,1}\right|}\,\,.\]
 Thus, the first regime corresponds to conditions where both nonlinear
crystals operate in the below threshold regime of parametric down-conversion.
Here the steady state solutions for the modes $\omega_{0}$, $\omega_{1}$
and $\omega_{2}$ are the same as in the usual nondegenerate parametric
down-conversion with a single crystal \cite{Reid-Drumm}. Figure~\ref{Fig: sss}
plots the steady state solution $n_{0}^{0}$ in the below threshold
regime where we have also introduced a new variable, namely $n_{0,cr}^{0}=|E_{thr,1}|^{2}/\gamma_{0}^{2}=\gamma^{2}/\chi^{2}$.
This is the critical value of $n_{0}^{0}$ at the first threshold.

\subsubsection{First above-threshold regime}

In the first above-threshold regime, the amplitudes of the modes $\omega_{3}$
and $\omega_{4}$ remain zero, while the amplitudes of the pump, signal
and idler modes ($\omega_{0}$, $\omega_{1}$ and $\omega_{2}$) are
nonzero. Accordingly, we again use the intensity and phase variables,
$n_{i}^{0}$ and $\phi_{i}^{0}$, $\alpha_{i}^{0}=\sqrt{n_{i}^{0}}\exp(i\phi_{i}^{0})$
for $i=0,1,2$, and write the steady state solutions as: \begin{align}
\alpha_{3}^{0} & =\alpha_{4}^{0}=0,\nonumber \\
n_{0}^{0} & =\frac{\gamma^{2}}{\chi^{2}},\nonumber \\
n_{1}^{0} & =n_{2}^{0}=\frac{\left|E_{0}\right|}{\chi}-\frac{\gamma_{0}\gamma}{\chi^{2}},\label{R2-intensities}\end{align}
 \begin{align}
\phi_{0}^{0} & =\varphi_{0},\nonumber \\
\phi_{1}^{0}+\phi_{2}^{0} & =\phi_{0}^{0}=\varphi_{0}.\label{R2-phases}\end{align}
 We see that the steady state intensities $n_{1}^{0}$ and $n_{2}^{0}$
correspond to physical solutions ($n_{i}^{0}>0$ ) if the driving
field intensity is above the first threshold, $\left|E_{0}\right|^{2}>\left|E_{thr,1}\right|^{2}$.
On the other hand, the linearized eigenvalue analysis for the sub-system
of intensity variable (see below) shows that the solutions are stable
for $\left|E_{0}\right|^{2}$ below a second threshold, $\left|E_{0}\right|^{2}<\left|E_{thr,2}\right|^{2}$,
where \begin{equation}
\left|E_{thr,2}\right|^{2}\equiv\frac{\gamma_{0}^{2}\gamma^{2}}{\chi^{2}}\left(1+\frac{\gamma}{\gamma_{0}}\right)^{2}=\left|E_{thr,1}\right|^{2}\left(1+\frac{\gamma}{\gamma_{0}}\right)^{2}.\label{threshold2}\end{equation}
 This implies that the first above threshold regime is restricted
to: \begin{equation}
\left|E_{thr,1}\right|^{2}<\left|E_{0}\right|^{2}<\left|E_{thr,2}\right|^{2}.\label{region2}\end{equation}
This is shown in Fig.~\ref{Fig: sss} along with the steady state
solutions of Eq.~(\ref{R2-intensities}).

In this regime, the first nonlinear crystal operates in the above-threshold
(stimulated) regime, while the operation of the second nonlinear crystal
is in the below-threshold (spontaneous) regime. The steady state solutions
for the $\omega_{0}$, $\omega_{1}$ and $\omega_{2}$ modes are the
same as in nondegenerate parametric down-conversion with a single
crystal \cite{Reid-Drumm}, except that the stability region has now
an upper bound.

\subsubsection{Second above-threshold regime}

In the second above-threshold regime, both nonlinear crystals operate
with stimulated emission, and the amplitudes of all intracavity modes
are nonzero. The mode $\omega_{2}$ acts as the pump mode with respect
to the second nonlinear crystal and its intensity is above the respective
threshold for stimulated down-conversion $\omega_{2}\rightarrow\omega_{3}+\omega_{4}$.
Note that $|E_{thr,2}|$ is very close to $|E_{thr,1}|$ in the case
of a strongly damped or nonresonant primary pump $\hat{a}_{0}$ that
we consider here. This makes this second above-threshold regime quite
accessible experimentally and, in the limit $\gamma_{2}\rightarrow0$,
could bring about effective $\x3$ behavior (see next section).

Again using the intensity and phase variables, the steady state solutions
can be written as follows: \begin{align}
 & n_{0}^{0}=n_{1}^{0}=\frac{\left|E_{0}\right|^{2}}{\left(\gamma_{0}+\gamma\right)^{2}},\nonumber \\
 & n_{2}^{0}=\frac{\gamma^{2}}{\chi^{2}},\nonumber \\
 & n_{3}^{0}=n_{4}^{0}=\frac{\left|E_{0}\right|^{2}}{\left(\gamma_{0}+\gamma\right)^{2}}-\frac{\gamma^{2}}{\chi^{2}},\label{R3-intensities}\end{align}
 \begin{align}
\phi_{0}^{0} & =\varphi_{0},\nonumber \\
\phi_{1}^{0}+\phi_{2}^{0} & =\phi_{0}^{0}=\varphi_{0},\nonumber \\
\phi_{3}^{0}+\phi_{4}^{0}-\phi_{2}^{0} & =0.\label{R3-phases}\end{align}
 The intensities $n_{1}^{0}$, $n_{2}^{0}$, and $n_{3}^{0}$ ($n_{3}^{0}=n_{4}^{0}$)
are related by a simple relationship \begin{equation}
n_{1}^{0}=n_{2}^{0}+n_{3}^{0}.\label{conservation}\end{equation}
 This reflects the photon number conservation in the second crystal
and the correlation between the photons $\omega_{1}$ and $\omega_{2}$,
including the possibility of conversion of photons $\omega_{2}$ into
a pair of photons $\omega_{3}$ and $\omega_{4}$. From the expressions
for $n_{3}^{0}$ and $n_{4}^{0}$, we see that physical solutions
are realized for driving field intensities above the second threshold,
\begin{equation}
\left|E_{0}\right|^{2}>\left|E_{thr,2}\right|^{2}.\label{region3}\end{equation}
 In addition, we show in the next section that the linearized eigenvalue
analysis reveals that the sub-system of intensity variables is stable
in this region. Thus, the second above-threshold regime corresponds
to Eq.~(\ref{region3}) and is pictured in Fig.~\ref{Fig: sss}
with its corresponding steady state solutions.

\section{Stability properties}

Here we give the details of the linearized eigenvalue analysis to
determine stability of the classical steady-state regimes. In order
to explain this approach, we proceed with a dimensionless analysis,
in terms of a small parameter \begin{equation}
g=\frac{\chi}{\gamma}\,\,.\end{equation}
 We now wish to derive the leading order behavior of the stochastic
fluctuations in each mode, as an expansion in terms of $g$. It is
simplest to first transform to dimensionless parameters, defining
dimensionless time as: \begin{align}
\tau=\gamma t\end{align}
 This scaled time variable will be used for all derivatives defined
in this section. Furthermore we will also use the dimensionless parameter
\begin{align}
\gamma_{r}=\frac{\gamma_{0}}{\gamma}\end{align}

We note here that a linearized analysis is only the first stage in
a stochastic diagram perturbation expansion\cite{Chaturvedi}, which
in general needs to be taken to higher order to reveal non-Gaussian
behavior\cite{drummond_nondeg}. The details of this will be treated
elsewhere.

\subsection{Positive-P method}

We start by using the full positive-P method to treat this system,
together with an appropriate scaling for the below threshold fields,
by introducing:\begin{eqnarray*}
\beta_{0} & = & \left(\alpha_{0}-\alpha_{0}^{0}\right)/g\\
\beta_{3} & = & \alpha_{3}/\sqrt{g}\\
\beta_{4} & = & \alpha_{4}/\sqrt{g}\end{eqnarray*}
 Using the semi-classical steady state solutions, Eq.~(\ref{eq of motion}),
and dropping higher-order terms of order $\sqrt{g}$ or higher, we
get: \begin{eqnarray*}
\dot{\beta}_{0} & = & -\gamma_{r}\beta_{0}-\alpha_{1}\alpha_{2}\\
\dot{\alpha}_{1} & = & -\alpha_{1}+\epsilon\alpha_{2}^{+}+\sqrt{\epsilon}\eta_{1}(\tau)\\
\dot{\alpha}_{2} & = & -\alpha_{2}+\epsilon\alpha_{1}^{+}+\sqrt{\epsilon}\eta_{2}(\tau)\\
\dot{\beta}_{3} & = & -\beta_{3}+\sqrt{\alpha_{2}}\eta_{3}(\tau)\\
\dot{\beta}_{4} & = & -\beta_{4}+\sqrt{\alpha_{2}}\eta_{4}(\tau)\end{eqnarray*}
 together with the Hermitian conjugate equations. The nonzero steady-state
correlations of the noise terms are: \begin{align}
\left\langle \eta_{1}(\tau)\eta_{2}(\tau')\right\rangle  & =\delta(\tau-\tau^{\prime}),\nonumber \\
\left\langle \eta_{3}(\tau)\eta_{4}(\tau')\right\rangle  & =\delta(\tau-\tau^{\prime}),\label{noises1}\end{align}
 The linearized equations for $\beta_{0}$ and $\beta{}_{3,4}$ are
all decoupled and have negative eigenvalues $-\gamma_{0}$ and $-\gamma$,
respectively. Accordingly, the corresponding steady states are stable.
The deterministic part of linearized equations for the remaining variables,
$\alpha_{1}$ and $\alpha_{2}$ (together with $\alpha_{1}^{+}$ and
$\alpha_{2}^{+}$), can be written in the matrix form as follows:
\begin{equation}
\underline{\dot{\alpha}}=A^{0}\underline{\alpha},\end{equation}
 where $\underline{\alpha}=\left(\alpha_{1},\alpha_{2},\alpha_{1}^{+},\alpha_{2}^{+}\right)^{T}$
and the drift matrix $A^{0}$ is given by \begin{equation}
A^{0}=\left(\begin{array}{cccc}
-1 & 0 & 0 & \epsilon\\
0 & -1 & \epsilon & 0\\
0 & \epsilon & -1 & 0\\
\epsilon & 0 & 0 & -1\end{array}\right).\end{equation}
 The eigenvalues of the matrix $A^{0}$ can be calculated explicitly,
with the result that their real parts are all negative if \begin{equation}
\epsilon^{2}<1.\end{equation}
 This defines the stability region, Eq.~(\ref{region1}), for the
steady states (\ref{SS1}) and the first threshold, Eq.~(\ref{threshold1}).

\subsection{Above-threshold stability}

In this section we analyze the stability of the above-threshold regimes.
For reasons of length, we do not give a complete analysis of the fluctuations,
but rather we simply determine which are the stable regimes. This
allows us to build a complete large-signal phase-diagram, which is
highly useful for determining the down-conversion properties of the
cascaded device. Detailed spectral properties will be analyzed elsewhere.

\subsubsection{First above-threshold regime}

Inspecting the semi-classical steady state solutions, Eqs. (\ref{R2-intensities})
and (\ref{R2-phases}), we immediately notice that while the sum of
the steady state phases $\phi_{1}^{0}+\phi_{2}^{0}$ of the signal
and idler modes is well defined and is equal to the phase of the driving
field, $\varphi_{0}$, the individual values of $\phi_{1}^{0}$ and
$\phi_{2}^{0}$ remain unknown. In other words, there is no unique
solution for the individual phases $\phi_{1}^{0}$ and $\phi_{2}^{0}$
and any attempt to perform linearization around any chosen value of
$\phi_{1}^{0}$ or $\phi_{2}^{0}$ will generate a zero eigenvalue,
implying that the steady states are unstable. This problem is known
as phase diffusion \cite{graham-phase-diff,Reid-Drumm}.

In order to correctly analyze the set of coupled equations of motion
in this regime, it is helpful to factorize them into a subset that
can be linearized and is stable, while the equation associated with
the zero eigenvalue must be isolated (decoupled) and treated exactly
without the use of linearization. This can be achieved by means of
transforming to a new set of stochastic variables. In doing so, we
note that the stochastic equations of motion for this system, Eqs.~(\ref{eq of motion}),
are equivalent in either It${\rm \hat{o}}$ or Stratonovich formulation
of the stochastic calculus. We employ the Stratonovich formulation
which has the advantage that the variable changes are achieved using
the usual calculus rules, without any extra variable-change terms.
Accordingly, we first transform to new intensity and phase variables
for the modes $\omega_{0}$, $\omega_{1}$, and $\omega_{2}$: \begin{align}
 & n_{j}=\alpha_{j}\alpha_{j}^{+},\nonumber \\
 & \phi_{j}=\frac{1}{2i}\ln\left(\frac{\alpha_{j}}{\alpha_{j}^{+}}\right),\;(j=0,1,2),\label{intensity-phase-vars}\end{align}
 which we note are complex. The stochastic variables $\alpha_{3,4}$
and $\alpha_{3,4}^{+}$, on the other hand, are transformed to: \begin{align}
\widetilde{\alpha}_{3,4} & =\alpha_{3,4}e^{-i\phi_{2}/2},\nonumber \\
\widetilde{\alpha}_{3,4}^{+} & =\alpha_{3,4}^{+}e^{i\phi_{2}/2}.\end{align}
 In these new variables, the stochastic differential equations become:
\begin{align}
\dot{n}_{0} & =-2\gamma_{0}n_{0}+2\left|E_{0}\right|\cos\left(\varphi_{0}-\phi_{0}\right)\nonumber \\
 & -2\chi\sqrt{n_{0}n_{1}n_{2}}\cos\left(\phi_{0}-\phi_{+}\right),\\
\dot{n}_{1} & =-2\gamma n_{1}+2\chi\sqrt{n_{0}n_{1}n_{2}}\cos\left(\phi_{0}-\phi_{+}\right)\nonumber \\
 & +F_{1}(t),\\
\dot{n}_{2} & =-2\gamma n_{2}+2\chi\sqrt{n_{0}n_{1}n_{2}}\cos\left(\phi_{0}-\phi_{+}\right)\nonumber \\
 & -\chi\sqrt{n_{2}}\left(\widetilde{\alpha}_{3}\widetilde{\alpha}_{4}+\widetilde{\alpha}_{3}^{+}\widetilde{\alpha}_{4}^{+}\right)+F_{2}(t),\label{SDE-intensities-R2}\end{align}
 \begin{align}
\dot{\phi}_{0} & =\frac{\left|E_{0}\right|}{n_{0}}\sin\left(\phi_{0}-\phi_{+}\right)\nonumber \\
 & \,\,\,\,-\chi\sqrt{\frac{n_{1}n_{2}}{n_{0}}}\sin\left(\phi_{0}-\phi_{+}\right),\\
\dot{\phi}_{1} & =\chi\sqrt{\frac{n_{0}n_{2}}{n_{1}}}\sin\left(\phi_{0}-\phi_{+}\right)+f_{1}(t),\\
\dot{\phi}_{2} & =\chi\sqrt{\frac{n_{0}n_{1}}{n_{2}}}\sin\left(\phi_{0}-\phi_{+}\right)\nonumber \\
 & -\frac{\chi}{2i\sqrt{n_{2}}}\left(\widetilde{\alpha}_{3}\widetilde{\alpha}_{4}-\widetilde{\alpha}_{3}^{+}\widetilde{\alpha}_{4}^{+}\right)+f_{2}(t),\label{SDE-phases-R2}\end{align}
 \begin{align}
\dot{\widetilde{\alpha}}_{3} & =-\gamma\widetilde{\alpha}_{3}+\chi\sqrt{n_{2}}\widetilde{\alpha}_{4}^{+}-\frac{i\chi}{2}\sqrt{\frac{n_{0}n_{1}}{n_{2}}}\widetilde{\alpha}_{3}\sin\left(\phi_{0}-\phi_{+}\right)\nonumber \\
 & +\frac{\chi}{4\sqrt{n_{2}}}\left(\widetilde{\alpha}_{3}\widetilde{\alpha}_{4}-\widetilde{\alpha}_{3}^{+}\widetilde{\alpha}_{4}^{+}\right)\widetilde{\alpha}_{3}+\mathcal{F}_{3}(t),\\
\dot{\widetilde{\alpha}}_{4} & =-\gamma\widetilde{\alpha}_{4}+\chi\sqrt{n_{2}}\widetilde{\alpha}_{3}^{+}-\frac{i\chi}{2}\sqrt{\frac{n_{0}n_{1}}{n_{2}}}\widetilde{\alpha}_{4}\sin\left(\phi_{0}-\phi_{+}\right)\nonumber \\
 & +\frac{\chi}{4\sqrt{n_{2}}}\left(\widetilde{\alpha}_{3}\widetilde{\alpha}_{4}-\widetilde{\alpha}_{3}^{+}\widetilde{\alpha}_{4}^{+}\right)\widetilde{\alpha}_{4}+\mathcal{F}_{4}(t).\label{tilde-alphas}\end{align}
 together with the equations for $\widetilde{\alpha}_{3}^{+}$ and
$\widetilde{\alpha}_{4}^{+}$. Here, we have defined the sum of the
phase variables $\phi_{1}$ and $\phi_{2}$ via \begin{equation}
\phi_{+}\equiv\phi_{1}+\phi_{2},\end{equation}
 and we note that the equations for $\widetilde{\alpha}_{3}$ and
$\widetilde{\alpha}_{4}$ contain terms that come from the time derivative
of $\phi_{2}$ which have been substituted with the right-hand side
of Eq.~(\ref{SDE-phases-R2}).

The new noise terms in the above set of equations of motion are defined
according to: \begin{align}
F_{1,2} & =\alpha_{1,2}^{+}\sqrt{\chi\alpha_{0}}\zeta_{1,2}+\alpha_{1,2}\sqrt{\chi\alpha_{0}^{+}}\zeta_{1,2}^{+},\\
f_{1,2} & =\frac{\sqrt{\chi\alpha_{0}}}{2i\alpha_{1,2}}\zeta_{1,2}-\frac{\sqrt{\chi\alpha_{0}^{+}}}{2i\alpha_{1,2}^{+}}\zeta_{1,2}^{+},\\
\mathcal{F}_{3,4} & =\sqrt{\chi\alpha_{2}}e^{-i\phi_{2}/2}\zeta_{3,4}-\frac{i\widetilde{\alpha}_{3,4}}{2}f_{2}.\end{align}
 These must be rewritten in terms of the intensity and phase variables
$n_{i}$ and $\phi_{i}$, for self-consistency: \begin{align}
F_{1,2} & =\sqrt{\chi n_{1,2}}(n_{0})^{1/4}\left[e^{-i\phi_{1,2}+i\phi_{0}/2}\zeta_{1,2}\right.\nonumber \\
 & \left.+e^{i\phi_{1,2}-i\phi_{0}/2}\zeta_{1,2}^{+}\right],\end{align}
 \begin{align}
f_{1,2} & =\frac{\sqrt{\chi}(n_{0})^{1/4}}{2i\sqrt{n_{1,2}}}\left[e^{-i\phi_{1,2}+i\phi_{0}/2}\zeta_{1,2}\right.\nonumber \\
 & \left.-e^{i\phi_{1,2}-i\phi_{0}/2}\zeta_{1,2}^{+}\right],\end{align}
 \begin{equation}
\mathcal{F}_{3,4}=\sqrt{\chi}(n_{2})^{1/4}\zeta_{3,4}-\frac{i\widetilde{\alpha}_{3,4}}{2}f_{2}.\;\end{equation}
 We next introduce the phase sum and difference variables, \begin{equation}
\phi_{\pm}=\phi_{1}\pm\phi_{2},\end{equation}
 and convert the equations of motion for $\phi_{1}$ and $\phi_{2}$
into: \begin{align}
\dot{\phi}_{+} & =\chi\sqrt{n_{0}}\left(\sqrt{\frac{n_{2}}{n_{1}}}+\sqrt{\frac{n_{1}}{n_{2}}}\right)\sin\left(\phi_{0}-\phi_{+}\right)\nonumber \\
 & -\frac{\chi}{2i\sqrt{n_{2}}}\left(\widetilde{\alpha}_{3}\widetilde{\alpha}_{4}-\widetilde{\alpha}_{3}^{+}\widetilde{\alpha}_{4}^{+}\right)+f_{+}(t),\\
\dot{\phi}_{-} & =\chi\sqrt{n_{0}}\left(\sqrt{\frac{n_{2}}{n_{1}}}-\sqrt{\frac{n_{1}}{n_{2}}}\right)\sin\left(\phi_{0}-\phi_{+}\right)\nonumber \\
 & +\frac{\chi}{2i\sqrt{n_{2}}}\left(\widetilde{\alpha}_{3}\widetilde{\alpha}_{4}-\widetilde{\alpha}_{3}^{+}\widetilde{\alpha}_{4}^{+}\right)+f_{-}(t),\label{phi-minus}\end{align}
 where the noise terms are \begin{equation}
f_{\pm}=f_{1}\pm f_{2}.\end{equation}

We now immediately see, that the equations of motion for the variables
$n_{0}$, $n_{1}$, $n_{2}$, $\widetilde{\alpha}_{3,4}$, $\widetilde{\alpha}_{3,4}^{+}$,
$\phi_{0}$ and $\phi_{+}$ are decoupled from the equation of motion
for the phase-difference variable $\phi_{-}$. All these variables
except $\phi_{-}$ have a unique semi-classical steady state solution
given by Eqs.~(\ref{R2-intensities})-(\ref{R2-phases}), with $\phi_{+}^{0}=\phi_{1}^{0}+\phi_{2}^{0}=\varphi_{0}$
and $\widetilde{\alpha}_{3,4}^{0}=0$ (along with $(\widetilde{\alpha}_{3,4}^{0})^{+}=(\widetilde{\alpha}_{3,4}^{0})^{\ast}=0$).
As we will show below, the linearized equations for this subsystem
of variables are stable, and therefore these variables can be treated
by means of linearization around their semi-classical steady states.
Indeed, by introducing small fluctuations around the steady states
\begin{align}
\delta n_{0,1,2}(t) & =n_{0,1,2}(t)-n_{0,1,2}^{0},\\
\delta\widetilde{\alpha}_{3,4}(t) & =\widetilde{\alpha}_{3,4}(t)-\widetilde{\alpha}_{3,4}^{0},\\
\delta\widetilde{\alpha}_{3,4}^{+}(t) & =\widetilde{\alpha}_{3,4}^{+}(t)-(\widetilde{\alpha}_{3,4}^{0})^{\ast},\\
\delta\phi_{0,+}(t) & =\phi_{0,+}(t)-\phi_{0,+}^{0},\end{align}
 we obtain the following set of linearized equations: \begin{align}
 & \delta\dot{n}_{0}=-\gamma_{0}\delta n_{0}-\gamma\delta n_{+},\label{R2-linear-n0}\\
 & \delta\dot{n}_{+}=\frac{2\chi n_{1}^{0}}{\gamma}\delta n_{0}+F_{+}^{0}(t),\\
 & \delta\dot{n}_{-}=-2\gamma\delta n_{-}+F_{-}^{0}(t),\end{align}
 \begin{align}
 & \delta\dot{\phi}_{0}=-\gamma_{0}\delta\phi_{0}-\frac{\chi^{2}n_{1}^{0}}{\gamma}\delta\phi_{+},\\
 & \delta\dot{\phi}_{+}=-2\chi\delta\phi_{+}+2\gamma\delta\phi_{0}+f_{+}^{0}(t),\end{align}
 \begin{align}
\delta\dot{\widetilde{\alpha}}_{3} & =-\gamma\delta\widetilde{\alpha}_{3}+\chi\sqrt{n_{1}^{0}}\delta\widetilde{\alpha}_{4}^{+}+\mathcal{F}_{3}^{0}(t),\\
\delta\dot{\widetilde{\alpha}}_{4} & =-\gamma\delta\widetilde{\alpha}_{4}+\chi\sqrt{n_{1}^{0}}\delta\widetilde{\alpha}_{3}^{+}+\mathcal{F}_{4}^{0}(t),\label{R2-linear-alpha4}\end{align}
 together with the equations for $\delta\widetilde{\alpha}_{3,4}^{+}$.
Here, we have used the explicit expression for the steady state solution
$n_{0}^{0}$ from Eq.~(\ref{R2-intensities}) and the fact that $n_{1}^{0}=n_{2}^{0}$.
The nonzero steady-state correlations of the noise terms, in the small-noise
approximation, are given by: \begin{align}
\left\langle F_{+}^{0}(t)F_{+}^{0}(t^{\prime})\right\rangle  & =-\left\langle F_{-}^{0}(t)F_{-}^{0}(t^{\prime})\right\rangle \nonumber \\
 & =4\gamma n_{1}^{0}\delta(t-t^{\prime}),\label{noise-F12}\\
\left\langle f_{+}^{0}(t)f_{+}^{0}(t^{\prime})\right\rangle  & =-\frac{\gamma}{n_{1}^{0}}\delta(t-t^{\prime}),\label{noise-f-plus}\\
\left\langle \mathcal{F}_{3}^{0}(t)\mathcal{F}_{4}^{0}(t^{\prime})\right\rangle  & =\chi\sqrt{n_{1}^{0}}\delta(t-t^{\prime}),\label{noise-F34}\end{align}
 By substituting the steady state intensity $n_{1}^{0}$ from Eqs.
(\ref{R2-intensities}), the linearized equations and hence their
solutions can be expressed in terms of the driving field intensity
$\left|E_{0}\right|^{2}$.

The eigenvalue analysis of the deterministic drift terms of the linearized
equations reveals that the equations for $\delta n_{0,+}$, $\delta n_{-}$,
and $\delta\phi_{0,+}$ are stable everywhere (the eigenvalues have
negative real parts), while the subsystem of variables $\left(\delta\widetilde{\alpha}_{3},\delta\widetilde{\alpha}_{4},\delta\widetilde{\alpha}_{3}^{+},\delta\widetilde{\alpha}_{4}^{+}\right)$
is stable only if \begin{equation}
\left|E_{0}\right|^{2}<\frac{\gamma_{0}^{2}\gamma^{2}}{\chi^{2}}\left(1+\frac{\gamma}{\gamma_{0}}\right)^{2}.\end{equation}
 This defines the second threshold, Eq.~(\ref{threshold2}), and
hence the upper bound on the driving field intensity $\left|E_{0}\right|^{2}$
for the first above-threshold region, Eq.~(\ref{region2}).

The remaining equation for the phase difference variable $\phi_{-}$,
Eq.~(\ref{phi-minus}), can not be linearized since the steady state
solution is not well defined and linearization around any chosen value
$\phi_{-}^{0}$ will reveal a zero eigenvalue, implying that the equation
is not stable. The right hand side of Eq.~(\ref{phi-minus}) can,
however, be simplified since all variables here can be linearized
around their stable steady states. Thus, expanding these in terms
of the stable steady states plus small fluctuations and keeping only
the linear terms, we see that the deterministic terms all cancel each
other. The resulting equation is \begin{equation}
\dot{\phi}_{-}=f_{-}^{0}(t),\label{phase-diffuse}\end{equation}
 with the following nonzero correlation of the noise term: \begin{equation}
\left\langle f_{-}^{0}(t)f_{-}^{0}(t^{\prime})\right\rangle =\frac{\gamma}{n_{1}^{0}}\delta(t-t^{\prime}).\end{equation}
 Thus, we have isolated the instability associated with a zero eigenvalue
into a single phase variable, which is the phase difference $\phi_{-}$
between the signal and idler phases. Unlike the other variables, the
phase difference $\phi_{-}$ is not a small fluctuation around a stable
steady state. Instead it undergoes continuous phase diffusion, governed
by the noise term $f_{-}^{0}(t)$ in Eq.~(\ref{phase-diffuse}).

Despite the fact that the noise terms $F_{1,2}$ and $f_{1,2}$ (and
hence $F_{+,-}\,\,$and$\,\, f_{+,-}$) depend explicitly on the individual
phases of the signal and idler modes, (which are not well-defined),
nevertheless, upon calculating the steady state noise correlations,
Eqs.~(\ref{noise-F12}) - (\ref{noise-F34}), these phases combine
into the phase sum $\phi_{+}=\phi_{1}+$ $\phi_{2}$ which has a well
defined steady state value and is stable. As a result, calculation
of observables via the solutions of the linearized equations of motion,
Eqs.~(\ref{R2-linear-n0} ) - (\ref{R2-linear-alpha4}) which ultimately
depend on the noise correlations -- is a well defined procedure, and
is independent on the individual phases $\phi_{1}$ and $\phi_{2}$
.

\subsection{Second above-threshold regime}

In the second above threshold regime, both parametric down-converters
operate in the above-threshold regime. In addition to the phase diffusion
in the signal and idler modes $\omega_{1}$ and $\omega_{2}$, we
now have a second source of instability which comes from the phase
diffusion in the secondary signal-idler modes, $\omega_{3}$ and $\omega_{4}$.
To simplify our analysis, we assume here that the damping constant
of the pump mode $\gamma_{0}$ is much larger than the damping constants
of all the other modes, \begin{equation}
\gamma_{0}\gg\gamma.\label{adiabatic}\end{equation}
 Under this condition, one can adiabatically eliminate the pump mode
from the equations of motion, Eq.~(\ref{eq of motion}), and restrict
ourselves to the dynamics of the remaining modes $\omega_{1}$, $\omega_{2}$,
$\omega_{3}$, and $\omega_{4}$. Thus, we assume that $\dot{\alpha}_{0}=0$
during the evolution of the amplitudes $\alpha_{1,2,3,4}$, and we
use the resulting expression for $\alpha_{0}$, \begin{equation}
\alpha_{0}=\frac{1}{\gamma_{0}}\left(E_{0}-\chi\alpha_{1}\alpha_{2}\right),\label{adiabatic-sol}\end{equation}
 (together with the expression for $\alpha_{0}^{+}$) in the equations
for $\alpha_{1,2,3,4}$. Transforming then to the intensity and phase
variables, as in Eq.~(\ref{intensity-phase-vars}), we obtain the
following set of stochastic equations for the intensities: \begin{align}
\dot{n}_{1} & =-2\gamma n_{1}+\frac{2\chi\left|E_{0}\right|}{\gamma_{0}}\sqrt{n_{1}n_{2}}\cos\theta_{1}-\frac{2\chi^{2}}{\gamma_{0}}n_{1}n_{2}\nonumber \\
 & +F_{1}(t),\label{R3-intensities-1}\\
\dot{n}_{2} & =-2\gamma n_{2}\,+\frac{2\chi\left|E_{0}\right|}{\gamma_{0}}\sqrt{n_{1}n_{2}}\cos\theta_{1}-\frac{2\chi^{2}}{\gamma_{0}}n_{1}n_{2}\nonumber \\
 & -2\chi\sqrt{n_{2}n_{3}n_{4}}\cos\theta_{2}+F_{2}(t),\\
\nonumber \\\dot{n}_{3} & =-2\gamma n_{3}+2\chi\sqrt{n_{2}n_{3}n_{4}}\cos\theta_{2}+F_{3}(t),\\
\nonumber \\\dot{n}_{4} & =-2\gamma n_{4}+2\chi\sqrt{n_{2}n_{3}n_{4}}\cos\theta_{2}+F_{4}(t),\label{R3-intensities-4}\end{align}
 Here, we have defined \begin{align}
\theta_{1} & =\phi_{1}+\phi_{2}-\varphi_{0},\\
\theta_{2} & =\phi_{3}+\phi_{4}-\phi_{2}.\end{align}
 which can serve as a new pair of phase variables, traded in favor
of the the signal and idler phases $\phi_{1}$ and $\phi_{2}$.

The stochastic equations of motion for the phase variables, which
we write at once in terms of $\theta_{1}$, $\theta_{2}$, $\phi_{3}$
and $\phi_{4}$, are: \begin{align}
\dot{\theta}_{1} & =-\frac{\chi\left|E_{0}\right|}{\gamma_{0}}\left(\sqrt{\frac{n_{1}}{n_{2}}}+\sqrt{\frac{n_{2}}{n_{1}}}\right)\sin\theta_{1}\nonumber \\
 & -\chi\sqrt{\frac{n_{3}n_{4}}{n_{2}}}\sin\theta_{2}+f_{\theta_{1}}(t),\label{R3-theta1}\\
\dot{\theta}_{2} & =-\chi\left(\sqrt{\frac{n_{2}n_{3}}{n_{4}}}+\sqrt{\frac{n_{2}n_{4}}{n_{3}}}-\sqrt{\frac{n_{3}n_{4}}{n_{2}}}\right)\sin\theta_{2}\nonumber \\
 & +\frac{\chi\left|E_{0}\right|}{\gamma_{0}}\sqrt{\frac{n_{1}}{n_{2}}}\sin\theta_{1}+f_{\theta_{2}}(t),\\
\dot{\phi}_{3} & =-\chi\sqrt{\frac{n_{2}n_{4}}{n_{3}}}\sin\theta_{2}+f_{3}(t),\label{R3-phi3}\\
\dot{\phi}_{4} & =-\chi\sqrt{\frac{n_{2}n_{3}}{n_{4}}}\sin\theta_{2}+f_{4}(t).\label{R3-phi4}\end{align}
 In the above equations, the noise terms are given by \begin{align}
f_{\theta_{1}} & =f_{1}+f_{2},\\
f_{\theta_{2}} & =f_{3}+f_{4}-f_{2},\end{align}
 and \begin{align}
f_{1,2} & =\frac{1}{2i\alpha_{1,2}}\sqrt{\frac{\chi}{\gamma_{0}}(E_{0}-\chi\alpha_{1}\alpha_{2})}\zeta_{1,2}\nonumber \\
 & -\frac{1}{2i\alpha_{j}^{+}}\sqrt{\frac{\chi}{\gamma_{0}}(E_{0}^{\ast}-\chi\alpha_{1}^{+}\alpha_{2}^{+})}\zeta_{1,2}^{+},\\
f_{3,4} & =\frac{\sqrt{\chi\alpha_{2}}}{2i\alpha_{3,4}}\zeta_{3,4}-\frac{\sqrt{\chi\alpha_{2}^{+}}}{2i\alpha_{3,4}^{+}}\zeta_{3,4}^{+}.\end{align}
 In addition, the noise terms $F_{1,2,3,4}$ in Eqs.~(\ref{R3-intensities-1}
)-(\ref{R3-intensities-4}) are given by \begin{align}
F_{1,2}(t) & =\alpha_{1,2}^{+}\sqrt{\frac{\chi}{\gamma_{0}}(E_{0}-\chi\alpha_{1}\alpha_{2})}\zeta_{1,2}\,\nonumber \\
 & +\alpha_{1,2}\sqrt{\frac{\chi}{\gamma_{0}}(E_{0}^{\ast}-\chi\alpha_{1}^{+}\alpha_{2}^{+})}\zeta_{1,2}^{+},\\
F_{3,4}(t) & =\alpha_{3,4}^{+}\sqrt{\chi\alpha_{2}^{{}}}\zeta_{3,4}+\alpha_{3,4}\sqrt{\chi\alpha_{2}^{+}}\zeta_{3,4}^{+}.\end{align}
 In all these noise terms the amplitude variables have to be expressed
in terms of the intensity and phase variables for self-consistency.

By inspecting Eqs.~(\ref{R3-intensities-1})-(\ref{R3-intensities-4})
and Eqs.~(\ref{R3-theta1})-(\ref{R3-phi4}), we see that the equations
for the intensities $n_{1,2,3,4}$ and phases $\theta_{1,2}$ are
decoupled from the equations for the phase variables, $\phi_{3,4}$.
The variables $n_{1,2,3,4}$ and $\theta_{1,2}$ all have well defined
semi-classical steady states, c.f., Eqs.~(\ref{R3-intensities})
and (\ref{R3-phases}), with $\theta_{1,2}^{0}=0$, and as we will
show below, the linearized eigenvalue analysis indicates their stability.
Thus, this subsystem of variables can be treated within the linearized
treatment of fluctuations. The phase variables $\phi_{3}$ and $\phi_{4}$,
on the other hand, do not have stable semi-classical steady states
and can not be treated by means of linearization. To demonstrate the
stability of the intensities $n_{1,2,3,4}$ and phases $\theta_{1,2}$,
we introduce fluctuations around the semi-classical steady states,
\begin{align}
\delta n_{j}(t) & =n_{j}(t)-n_{j}^{0},\;(j=1,2,3,4),\\
\delta\theta_{1,2}(t) & =\theta_{1,2}(t)-\theta_{1,2}^{0},\end{align}
 and derive the following linearized equations for the intensity fluctuations:
\begin{align}
\dot{\delta n}_{1} & =-\gamma\left(1+\frac{\gamma}{\gamma_{0}}\right)\delta n_{1}\nonumber \\
 & +\frac{\chi^{2}n_{1}^{0}}{\gamma}\left(1-\frac{\gamma}{\gamma_{0}}\right)\delta n_{2}+F_{1}^{0}(t),\label{R3-linearized-n1}\\
\dot{\delta n_{2}} & =-\gamma\left(1+\frac{\chi^{2}}{\gamma_{0}\gamma}\right)\delta n_{2}\nonumber \\
 & +\gamma\left(1-\frac{\gamma}{\gamma_{0}}\right)\delta n_{1}-\gamma\delta n_{+}+F_{2}^{0}(t),\\
\dot{\delta n_{+}} & =\frac{2\chi^{2}n_{3}^{0}}{\gamma}\delta n_{2}+F_{+}^{0}(t),\\
\dot{\delta n_{-}} & =-2\gamma\delta n_{-}+F_{-}^{0}(t),\label{R3-linearized-n-minus}\end{align}
 where we have additionally transformed to the intensity sum and difference
variables \begin{align}
 & \delta n_{+}=\delta n_{3}+\delta n_{4},\nonumber \\
 & \delta n_{-}=\delta n_{3}-\delta n_{4},\end{align}
 to further simplify the eigenvalue analysis. We have also defined
\begin{equation}
F_{\pm}^{0}\equiv F_{3}^{0}\pm F_{4}^{0}.\end{equation}
 As we see, the equation for the intensity difference $\delta n_{-}$
fluctuation is decoupled and immediately results in a negative eigenvalue
in the drift term, implying stability. The coupled equations for fluctuations
in $\delta n_{1}$, $\delta n_{2}$, and $\delta n_{+}$ result in
a cubic equation for the eigenvalues of the respective drift matrix.
While this cannot be solved explicitly, however, the negative real
parts of the eigenvalues required for stability is ascertained here
using the Routh-Hurwitz criterion \cite{mathhandbook}.

The linearized equations for the phase fluctuations $\delta\theta_{1,2}(t)$
are: \begin{align}
\dot{\delta\theta_{1}} & =-\frac{\chi\left|E_{0}\right|}{\gamma_{0}}\left(\sqrt{\frac{n_{1}^{0}}{n_{2}^{0}}}+\sqrt{\frac{n_{2}^{0}}{n_{1}^{0}}}\right)\delta\theta_{1}\nonumber \\
 & -\chi\frac{n_{3}^{0}}{\sqrt{n_{2}^{0}}}\delta\theta_{2}+f_{\theta_{1}}^{0},\label{R3-linearized-theta1}\\
\dot{\delta\theta_{2}} & =-\chi\left(2\sqrt{n_{2}^{0}}-\frac{n_{3}^{0}}{\sqrt{n_{2}^{0}}}\right)\delta\theta_{2}\nonumber \\
 & \,\,\,\,+\frac{\chi\left|E_{0}\right|}{\gamma_{0}}\sqrt{\frac{n_{1}^{0}}{n_{2}^{0}}}\delta\theta_{1}+f_{\theta_{2}}^{0},\label{R3-linearized-theta2}\end{align}
 The eigenvalues of the corresponding drift matrix can be found explicitly,
with the result that they all have negative real parts and therefore
the equations are stable. The nonzero steady state correlations of
the noise terms in the linearized Eqs.~(\ref{R3-linearized-n1})-(\ref{R3-linearized-n-minus})
and Eqs.~(\ref{R3-linearized-theta1})-(\ref{R3-linearized-theta2})
are: \begin{align}
\left\langle F_{1}^{0}(t)F_{2}^{0}(t^{\prime})\right\rangle  & =2\gamma n_{1}^{0}\delta(t-t^{\prime}),\\
\left\langle F_{+}^{0}(t)F_{+}^{0}(t^{\prime})\right\rangle  & =-\left\langle F_{-}^{0}(t)F_{-}^{0}(t^{\prime})\right\rangle \nonumber \\
 & =-2\left\langle F_{3}^{0}(t)F_{4}^{0}(t^{\prime})\right\rangle \nonumber \\
 & =4\gamma n_{3}^{0}\delta(t-t^{\prime}),\\
\left\langle f_{\theta_{1}}^{0}(t)f_{\theta_{1}}^{0}(t^{\prime})\right\rangle  & =-2\left\langle f_{\theta_{1}}^{0}(t)f_{\theta_{2}}^{0}(t^{\prime})\right\rangle \nonumber \\
 & =2\left\langle f_{1}^{0}(t)f_{2}^{0}(t^{\prime})\right\rangle \nonumber \\
 & =-\frac{\chi^{2}}{\gamma}\delta(t-t^{\prime}),\\
\left\langle f_{\theta_{2}}^{0}(t)f_{\theta_{2}}^{0}(t^{\prime})\right\rangle  & =2\left\langle f_{3}^{0}(t)f_{4}^{0}(t^{\prime})\right\rangle \nonumber \\
 & =-\frac{\gamma}{n_{3}^{0}}\delta(t-t^{\prime}).\label{f3f4}\end{align}

The remaining phase variables, $\phi_{3}$ and $\phi_{4}$, cannot
be treated within the linearized fluctuation treatment, however, the
right hand sides of the corresponding equations of motion, Eqs.~(\ref{R3-phi3})
and (\ref{R3-phi4} ), can be simplified since all variables here
have stable steady states and can be linearized. This gives \begin{align}
\dot{\phi_{3}} & =-\gamma\delta\theta_{2}+f_{3}^{0}(t),\label{R3-phi3-lin}\\
\dot{\phi_{4}} & =-\gamma\delta\theta_{2}+f_{4}^{0}(t).\label{R3-phi4-lin}\end{align}
 where the nonzero steady state correlation of the noise terms is
given in Eq.~(\ref{f3f4}). To further simplify the analysis we introduce
the sum and difference phase variables \begin{equation}
\theta_{\pm}=\phi_{3}\pm\phi_{4},\end{equation}
 for which the equations of motions are \begin{align}
\dot{\theta}_{+} & =-2\gamma\delta\theta_{2}+f_{\theta_{+}}^{0}(t),\label{theta-plus}\\
\dot{\theta}_{-} & =f_{\theta_{-}}^{0}(t),\label{theta-minus}\end{align}
 The source of instability for the phase variable $\theta_{-}$ is
obvious, while for the phase variable $\theta_{+}$ the presence of
a zero eigenvalue is revealed when the corresponding linearized equation
is combined with Eqs.~(\ref{R3-linearized-theta1})-(\ref{R3-linearized-theta2}).
Thus the variables $\theta_{+}$ and $\theta_{-}$ can not be linearized,
and have to be treated exactly. The nonzero correlations of the noise
terms $f_{\theta_{+}}^{0}=f_{3}^{0}+f_{4}^{0}$ and $f_{\theta_{-}}^{0}=f_{3}^{0}-f_{4}^{0}$
are \begin{align}
\left\langle f_{\theta_{+}}^{0}(t)f_{\theta_{+}}^{0}(t^{\prime})\right\rangle  & =-\left\langle f_{\theta_{-}}^{0}(t)f_{\theta_{-}}^{0}(t^{\prime})\right\rangle \nonumber \\
 & =2\left\langle f_{3}^{0}(t)f_{4}^{0}(t^{\prime})\right\rangle \nonumber \\
 & =-\frac{\gamma}{n_{3}^{0}}\delta(t-t^{\prime}).\end{align}
 From Eq.~(\ref{theta-plus}) we see that the dynamics of the phase
variable $\theta_{+}$ depends on phase fluctuations in $\delta\theta_{2}$,
and therefore the equation for $\theta_{+}$ has to be integrated
after solving for $\delta\theta_{2}$, Eqs.~(\ref{R3-linearized-theta1}
)-(\ref{R3-linearized-theta2}). The solution for $\theta_{+}(t)$
can be written as \begin{equation}
\theta_{+}(t)=\theta_{+}(t_{0})+\int_{t_{0}}^{t}\left[-2\gamma\delta\theta_{2}(t^{\prime})+f_{\theta_{+}}^{0}(t^{\prime})\right]dt^{\prime},\end{equation}
 while the solution for $\theta_{-}(t)$ is \begin{equation}
\theta_{-}(t)=\theta_{-}(t_{0})+\int_{t_{0}}^{t}f_{\theta_{-}}^{0}(t^{\prime})dt^{\prime}.\end{equation}
 Since $\delta\theta_{2}(t)$ as a solution to the set of linearized
Eqs.~(\ref{R3-linearized-theta1})-(\ref{R3-linearized-theta2})
depends on the noise terms $f_{\theta_{1}}^{0}$ and $f_{\theta_{2}}^{0}$,
the calculation of correlations involving the phase sum variable $\theta_{+}(t)$
will also depend on the following nonzero noise correlation: \begin{equation}
\left\langle f_{\theta_{+}}^{0}(t)f_{\theta_{2}}^{0}(t^{\prime})\right\rangle =2\left\langle f_{3}^{0}(t)f_{4}^{0}(t^{\prime})\right\rangle =-\frac{\gamma}{n_{3}^{0}}\delta(t-t^{\prime}),\end{equation}
 while $\left\langle f_{\theta_{+}}^{0}(t)f_{\theta_{1}}^{0}(t^{\prime})\right\rangle =0$. 

This completes the analysis of the system in the second above-threshold
regime.

\section{Numerical simulations}

A qualitative reasoning identifies the far-above-second-threshold
situation as interesting for mimicking a $\x3$ OPO, in the regime
where losses for the intermediate pump $\hat{a}_{2}$ are negligible,
i.e., $\gamma_{2}\ll\gamma_{1}=\gamma_{3}=\gamma_{4}\ll\gamma_{0}$
(note that this is different to the condition given in Eq.~(\ref{gammazero})).
Indeed, these hypotheses should yield close-to-ideal down-conversion
rate from field $\hat{a}_{2}$ to signal fields $\hat{a}_{3}$ and
$\hat{a}_{4}$, comparable to the emission rate into $\hat{a}_{1}$,
and therefore be consistent with the expectation of threefold quantum
correlations between $\hat{a}_{1}$, $\hat{a}_{3}$, $\hat{a}_{4}$,
which should be non-Gaussian (another favorable situation for this
effect would be the case $\chi_{2}\gg\chi_{1}$). The goal of the
following numerical simulations is therefore to ascertain the stability
of the resonant cascade in such cases, which are not covered by the
previous analytical treatment.

The numerical treatment is limited to the classical equations of motion,
given by Eq.~(\ref{eq of motion}) with $\zeta_{i}=0,i\in[0,4]$,
which are integrated numerically using a fourth order Runge-Kutta
routine similar to the method given in \cite{lugiato}. The first
and second nonlinearities were taken to be equal, i.e., $\chi_{1}=\chi_{2}$.
All down-converted fields were given minute initial amplitudes and
random initial phases, of which the subsequent dynamical phases were
independent.

\begin{figure}[htb]
\begin{centering}
\includegraphics[width=3.25in]{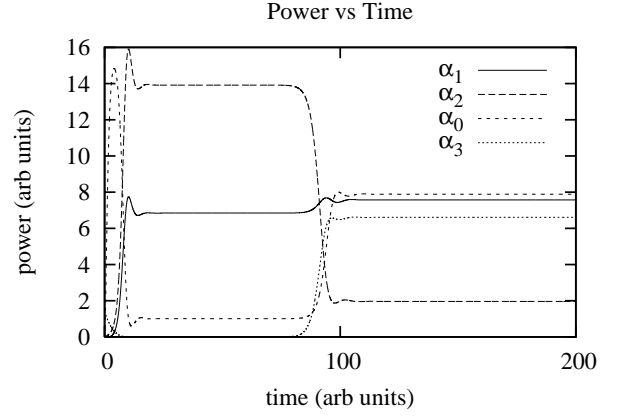} 
\par\end{centering}

\caption{OPO intracavity powers for $\gamma_{2}=0.08,\gamma_{1,3,4}=0.14,\gamma_{0}=2.0,|E_{0}|/|E_{thresh,2}|=3.5$,
for zero detunings. }

\label{fig:power_1std} 
\end{figure}

\begin{figure}[htb]
\begin{centering}
\includegraphics[width=3.25in]{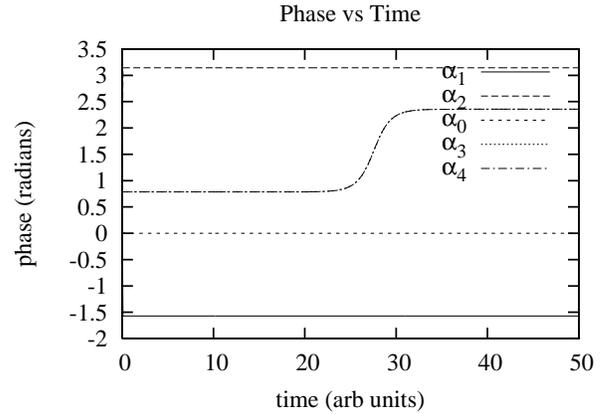} 
\par\end{centering}

\caption{Individual field phases, going through the second threshold, for $\gamma_{2}=0.08,\gamma_{1,3,4}=0.14,\gamma_{3}=1.0,|E_{0}|/|E_{thresh,2}|=3.5$,
for zero detunings.}

\label{fig:all_phase_thresh} 
\end{figure}

\begin{figure}[htb]
\begin{centering}
\includegraphics[width=3.25in]{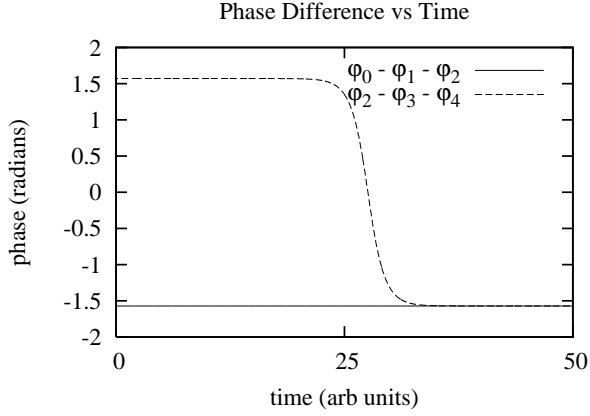} 
\par\end{centering}

\caption{The phase differences $\theta_{1}$ and $\theta_{2}$ between the
fields involved in the first and second stages of the nondegenerate
interaction, going through the second threshold, for $\gamma_{2}=0.08,\gamma_{1,3,4}=0.14,\gamma_{3}=1.0,|E_{0}|/|E_{thresh,2}|=3.5$,
for zero detunings. The stationary phase differences above the second
threshold indicate that down-conversion is taking place in both stages.}

\label{fig:phase_sum_thresh} 
\end{figure}

\begin{figure}[htb]
\begin{centering}
\includegraphics[width=3.25in]{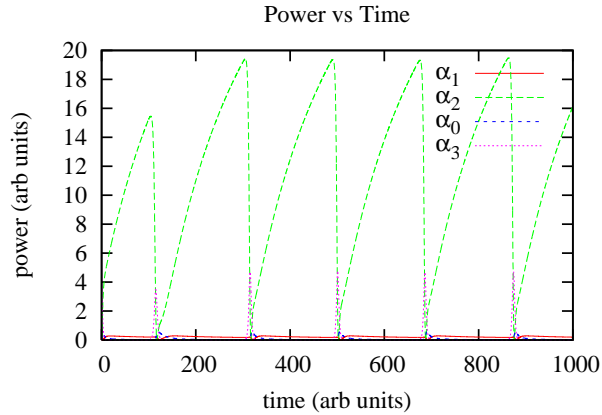} 
\par\end{centering}

\caption{OPO intracavity powers for $\gamma_{2}=0.001,\gamma_{1,3,4}=0.14,\gamma_{0}=1.0,|E_{0}|/|E_{thresh,2}|=5.8$,
for zero detunings. Note that $\alpha_{4}$ is omitted since its plot
follows exactly $\alpha_{3}$. The spiking frequency increases with
$|E_{0}|/|E_{thresh,2}|$ when other parameters are held constant.}

\label{fig:power_1} 
\end{figure}

\begin{figure}[htb]
\begin{centering}
\includegraphics[width=3.25in]{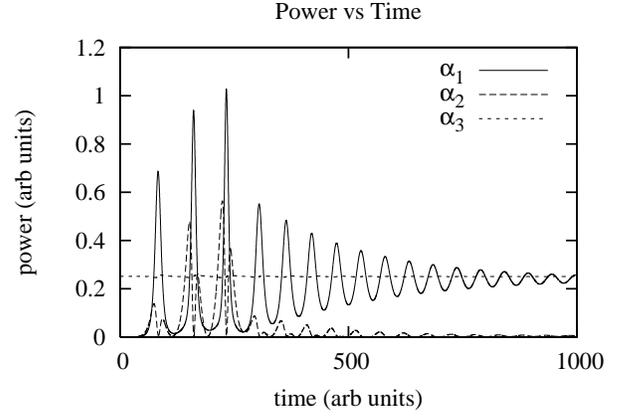} \includegraphics[width=3.25in]{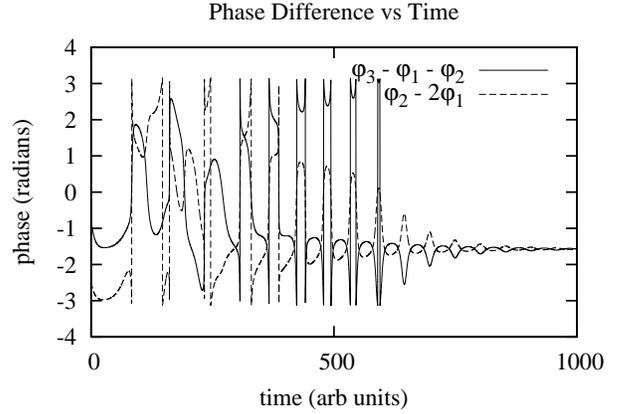} 
\par\end{centering}

\caption{Top: OPO intracavity powers for $(\gamma_{0},\gamma_{2},\gamma_{1})=(10,0,0.02)$,
$|E_{0}|/|E_{thresh,2}|=0.7$. Bottom: Phase differences $\theta_{1}$
and $\theta_{2}$, for the same conditions (phases are numerically
wrapped inside $[-\pi,\pi]$).}

\label{RCPthesis} 
\end{figure}

\begin{figure}[htb]
\begin{centering}
\includegraphics[width=3.25in]{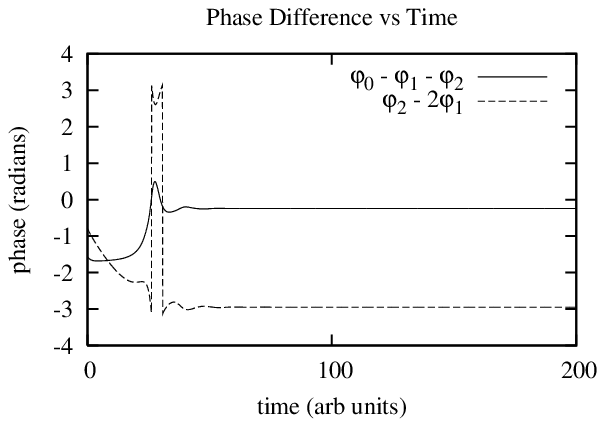} \includegraphics[width=3.25in]{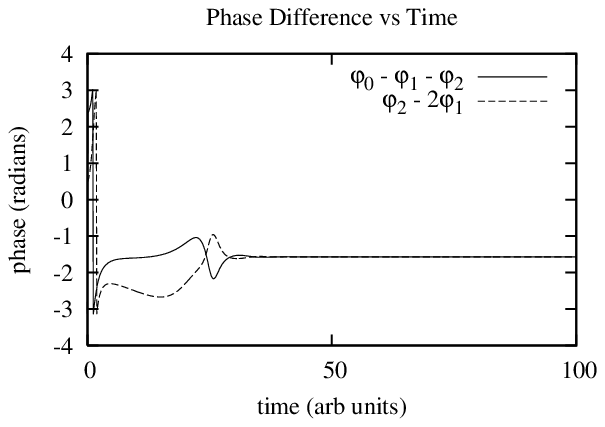} 
\par\end{centering}

\caption{Phase differences $\theta_{1}$ and $\theta_{2}$, for zero detunings
and $|E_{0}|/|E_{thresh,2}|=2.0$. Top: $\gamma_{2}=0.08$, $\gamma_{1}=0.14$.
Bottom: $\gamma_{1}=\gamma_{2}=0.14$ (phases are numerically wrapped
inside $[-\pi,\pi]$).}

\label{fig:degphasediff} 
\end{figure}

\begin{figure}[htb]
\begin{centering}
\includegraphics[width=3.25in]{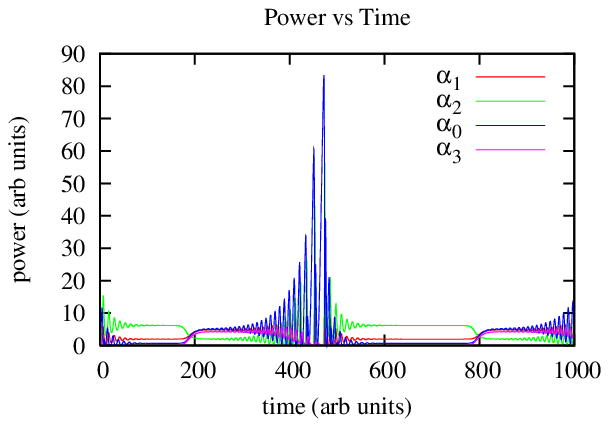} \includegraphics[width=3.25in]{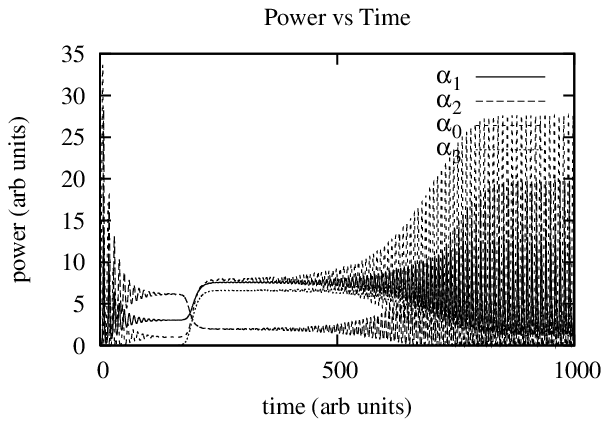} 
\par\end{centering}

\caption{OPO intracavity powers for $\gamma_{0}=0.08,\gamma_{1,3,4}=0.14,|E_{0}|/|E_{thresh,2}|=2.8$,
for zero detunings. Top: $\gamma_{2}=0.09$. Bottom: $\gamma_{2}=0.14$.}

\label{fig:cavityswitching} 
\end{figure}

\subsection{Steady-state solutions}

As mentioned above, we restrict our analysis to a set of parameters
such that $\gamma_{2}\ll\gamma_{1}=\gamma_{3}=\gamma_{4}\ll\gamma_{0}$.
That is, the primary pump mode is not resonant, the intermediate pump
is highly resonant (most of its losses occur in down-conversion),
and the signal fields are sufficiently resonant to acquire a threshold
as low as a typical single-stage, doubly resonant OPO (DRO). In this
case the OPO fields show decaying oscillations to a steady state after
reaching the second threshold. Higher losses for $\gamma_{2}$ result
in over-damping. Figure \ref{fig:power_1std} illustrates the steady-state
solutions. Both thresholds are clearly visible. At long times ($t>100$
in Fig.~\ref{fig:power_1std}) the field amplitudes match the stationary
solutions of the previous section. Increasing $E_{0}$ or $\gamma_{2}$
causes the oscillations to decay more quickly. 

Figure \ref{fig:all_phase_thresh} shows the individual phases as
the second threshold is reached. The final phase in the steady state
is independent of the initial starting phase of any of the fields'
seed values. (The primary pump parameter $E_{0}$ is taken to be real.)
Figure \ref{fig:phase_sum_thresh} shows the nonlinear phase differences
$\theta_{1}$ and $\theta_{2}$ for the first and second stages of
the OPO. It is clear from these phase differences that the system
is in a state of cascaded parametric down-conversion.

\begin{figure}[htbf]

\begin{centering}
\includegraphics[width=3.25in]{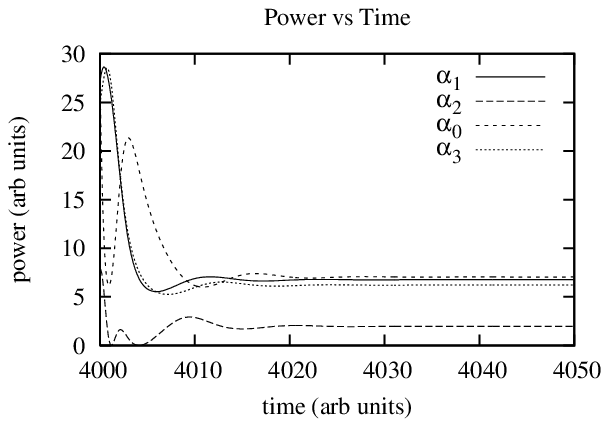} 
\par\end{centering}

\caption{Evolution of intracavity fields for $\gamma_{0}=1.0,\gamma_{2}=0.08,\gamma_{1,3,4}=0.14,|E_{0}|/|E_{thresh,2}|=3.5$,
perturbation $\delta_{\alpha i}=\alpha_{i}$, for zero detunings.
The perturbation has been applied simultaneously to the real and imaginary
parts of the fields, leaving no net perturbation to the phases.}

\label{fig:stabilityres} 
\end{figure}

\begin{figure}[htb]

\begin{centering}
\includegraphics[width=3.25in]{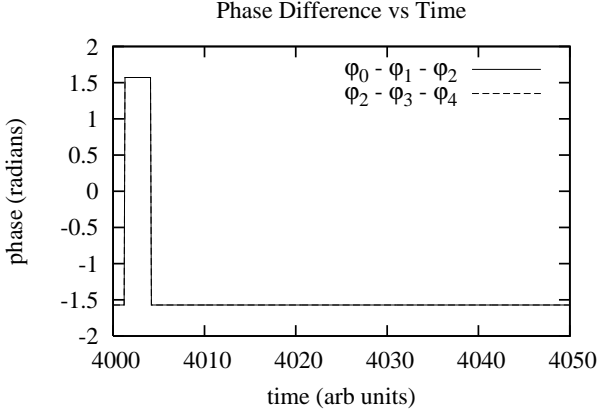} 
\par\end{centering}

\caption{Evolution of the phase differences $\theta_{1}$ and $\theta_{2}$
for $\gamma_{0}=1.0,\gamma_{2}=0.08,\gamma_{1,3,4}=0.14,|E_{0}|/|E_{thresh,2}|=3.5$,
perturbation $\delta_{\alpha i}=\alpha_{i}$, for zero detunings.
The perturbation has been applied to the imaginary parts of the fields.}

\label{fig:phaseresE2} 
\end{figure}

\begin{figure}[htb]

\begin{centering}
\includegraphics[width=3.25in]{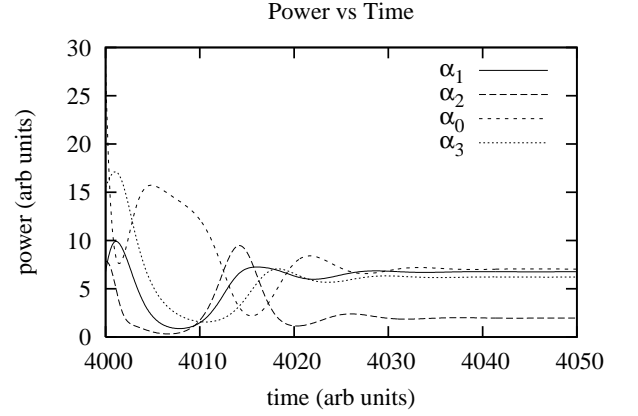} 
\par\end{centering}

\caption{Evolution of intracavity fields for $\gamma_{0}=1.0,\gamma_{2}=0.08,\gamma_{1,3,4}=0.14,|E_{0}|/|E_{thresh,2}|=3.5$,
perturbation $\delta_{\alpha i}=\alpha_{i}$, for zero detunings.
The perturbation has been applied to the real part of the field amplitudes.}

\label{fig:stabilityresphschg} 
\end{figure}

\begin{figure}[htb]

\begin{centering}
\includegraphics[width=3.25in]{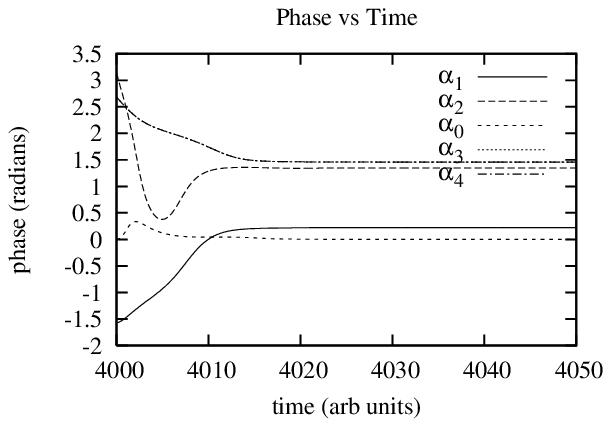} \includegraphics[width=3.25in]{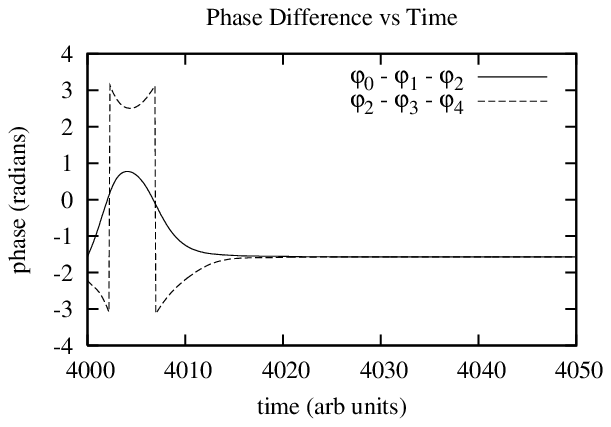} 
\par\end{centering}

\caption{Evolution of the OPO phases for $\gamma_{0}=1.0,\gamma_{2}=0.08,\gamma_{1,3,4}=0.14,|E_{0}|/|E_{thresh,2}|=3.5$,
perturbation $\delta_{\alpha i}=\alpha_{i}$, for zero detunings.
The perturbation has been applied to the real part of the fields.
Top: individual phases. Bottom: $\theta_{1}$ and $\theta_{2}$ (phases
are numerically wrapped inside $[-\pi,\pi]$).}

\label{fig:stabilitytotalresphschg} 
\end{figure}

\subsection{Second above-threshold regime for low $\gamma_{2}$}

\subsubsection{Nondegenerate cascade}

One interesting question is that of obtaining a stable effective $\x3$
OPO by lowering $\gamma_{2}$ and operating well above second threshold.
In that case, the nondegenerate and degenerate cascades do not exhibit
the same behavior. As the intermediate pump loss rate is lowered,
a spiking instability is obtained in both cases, as displayed in Fig.~\ref{fig:power_1}
for the nondegenerate case. One may overcome this self-pulsing and
induce a transition to a stable steady state by increasing the pump
parameter above threshold. The lower $\gamma_{2}$ the higher $|E_{0}|/|E_{thresh,2}|$
needs to be to achieve steady state in the nondegenerate case.

\subsubsection{Degenerate cascade}

In the degenerate case, a remarkable result is that the spiking behavior
is always transient and relaxes into a stationary state. However,
more insight into the behavior of the degenerate cascade is obtained,
once again, by scrutinizing the evolution of the phases $\theta_{1}$
and $\theta_{2}$. This bears particular physical significance for
the degenerate cascade because the first stage can stimulate emission
in the second stage, which cannot happen in the nondegenerate cascade
due to the indistinguishability of the signal fields \cite{note}.
Because of this effect, the degenerate cascade will exhibit greater
sensitivity to the evolution of $\theta_{1}$ and $\theta_{2}$, whose
swings translate into the appearance of competing sum-frequency generation
(SFG) processes in both stages.

\paragraph{First above-threshold:}

Figure \ref{RCPthesis} (top) displays the damping of the low $\gamma_{2}$
spiking. The phase behavior is plotted in Fig.~\ref{RCPthesis} (bottom)
and shows that significantly rich transient evolution eventually yields
a true cascade of two parametric down-conversions (PDCs). 

\paragraph{Second above-threshold:}

In Fig.~\ref{fig:degphasediff} (top), PDC is not the only process
occurring in both stages: the solutions can be seen to have a PDC
component and a competing SFG component. This is consistent with the
entering of the stimulated emission regime in the second stage as
one crosses the second threshold. The system is able to find a steady
state solution nonetheless. However, the quantum statistics might
be expected to be nontrivially affected. If one increases $\gamma_{2}$
to the level of the signal loss rate, the phase evolution yields this
time a stable PDC cascade (Fig.~\ref{fig:degphasediff} (bottom)),
as already demonstrated analytically in the previous section.

In conclusion, the degenerate cascade, because of the additional signal
feedback between the two stages, is clearly a much richer system than
the nondegenerate cascade. This additional feedback leads to stabilization
of the PDC cascade in the low $\gamma_{2}$ regime in the first above-threshold
regime. In the second above-threshold regime, however the degenerate
cascade displays two stable regimes, one of which does not have pure
PDC character. Bistable behavior or a bifurcation is to be expected
there. This also opens interesting horizons for the quantum simulations
of degenerate resonant cascades.

\subsection{Second above-threshold regime for low $\gamma_{2},\gamma_{0}$}

It is interesting to briefly investigate the behaviour found for low
$\gamma_{0}$. This regime involves a low-loss, resonant pump mode.
It has different stability properties to the situations treated elsewhere
in this paper, and more dramatic behavior is observed. By setting
$\gamma_{0}\sim\gamma_{2}<\gamma_{1,3,4}$, c.f. Fig.~\ref{fig:cavityswitching}
(top), the OPO becomes unstable above the second threshold, where
the pump $\hat{a}_{0}$ and intermediate pump $\hat{a}_{2}$ compete
strongly. The above-threshold phase leads to a return to the first-above
threshold regime, before recurring. In the case where $\gamma_{0}<\gamma_{2}$,
c.f., Fig.~\ref{fig:cavityswitching} (bottom), the amplitude of
the oscillations above the second threshold keep increasing and the
system never reverts to the first-above threshold regime.

\subsection{Stability analysis of stationary solutions}

We simulate the effect of a perturbation by causing an instantaneous
change in the intracavity fields and observing the numerical response
of the system. Of particular interest is the phase evolution of the
stationary solutions under two different types of perturbations, for
this will give insight into competing interaction (SFG/PDC) behaviors
in the degenerate case. We distinguish several types of perturbations:
{\em (i)} amplitude changes of the fields, leaving the phase unperturbed.
{\em (ii)} phase change of the fields. {\em (iii)} change in
both. We display typical results obtained for a variety of magnitudes
of change.

For small perturbations on the order of a couple percent of the steady
state amplitudes (perturbations to phase included), the system returns
to steady state after a few oscillations, and the field phases also
return to their steady state values. Some perturbations may change
the individual steady-state phases; however these changes are inconsequential
if the compound phase differences $\theta_{1}$ and $\theta_{2}$
remain at the PDC values. Figure \ref{fig:stabilityres} shows a typical
response of the intracavity powers to a perturbation on the field
amplitudes only. The OPO returns to the original steady state solutions
even under quite large disturbances. Figure \ref{fig:phaseresE2}
shows an important part of the phase evolution. Upon disturbing the
system, the phase of the secondary pump shifts by $\pi$ and then
back by $-\pi$. This is only true for a large change in the real
or imaginary components of the field (greater than $50\%$ in this
case). The phases of all the other fields remain comparatively unaffected.
Thus, the phase differences $\theta_{1}$ and $\theta_{2}$ shift
by $\pi$ quickly and then by $-\pi$ (see Fig.~\ref{fig:phaseresE2}).
When the disturbance is small $\sim10\%$, the phase differences recover
so quickly that a change in phase differences is not observed.

Figure \ref{fig:stabilityresphschg} shows a more complicated amplitude
response when a perturbation is applied to both the phase and amplitude
simultaneously. The OPO recovers the stationary amplitudes, but each
field, except the primary pump, also undergoes a permanent phase change,
c.f. Fig.~\ref{fig:stabilitytotalresphschg} (top), even though $\theta_{1}$
and $\theta_{2}$ return to unaltered values after opposite fluctuations,
c.f. Fig.~\ref{fig:stabilitytotalresphschg} (bottom). With increasing
pump parameter, the phase changes in Fig.~\ref{fig:stabilitytotalresphschg}
and Fig.~\ref{fig:phaseresE2}, can undergo several sign changes
before returning to steady state. This effect can be seen when the
applied perturbation is very large. These results indicate that nondegenerate
cascade is essentially as stable as a single-stage DRO.

\section{Conclusion}

In conclusion, we have given a preliminary analysis of the novel properties
of a doubly cascaded nondegenerate intracavity parametric oscillator.
This has the property that it is able to mimic a $\chi^{(3)}$ down-conversion
system, while still relying on the properties of widely available
phase-matched $\chi^{(2)}$ down-conversion crystals. Our analysis
focuses on constructing phase-space equations for the cascaded system,
and demonstrating the existence of multiple thresholds and stable
regions.

In the case of five non-degenerate modes, we have derived phase-space
equations in the full double-dimensional positive-P representation,
as well as approximate equations using the semi-classical or truncated
Wigner approach. We show the presence of three distinct classically
stable regimes, corresponding to below-threshold operation, an intermediate
threshold where only some of the modes are classically excited, and
a fully above threshold regime similar to $\chi^{(3)}$ down-conversion.

A detailed analysis of stability of these regimes is carried out to
show whether the relevant driving fields will result in stable operation.
This analysis is restricted to the non-degenerate case, for parameter
values in which all losses are equal except for the pump, which is
assumed to be strongly damped. The nonlinear coefficients are also
assumed to be equal. We find that for these parameter values each
of the three regimes mentioned is stable, that is, small signals are
damped back to the classical steady-state values.

We also give dynamical simulations of the mean field equations, which
clearly demonstrate the existence of stable regimes, as well as unusual
phase-evolution and distinct dynamical behaviour in the degenerate
and non-degenerate cases. Remarkable coincidences of two \cite{pfister}
and even three \cite{pooser} nonlinear interactions in a single-grating
periodically poled crystal have been observed, which illustrates the
experimental possibilities of such a technique. The dynamical analysis
in this case, although based on classical equations, is able to treat
a larger variety of parameters and detunings, as well as allowing
an investigation of stability in the case of much larger perturbations.
The general conclusion is that both the cascaded DPO and NDPO have
a rich variety of stable operating regimes and thresholds, including
the possibility of an above threshold $\chi^{(3)}$ domain.

CW, BP, KVK, and PDD acknowledge an Australian Research Council Centre
of Excellence grant for the support of this work. RCP and OP acknowledge
support by NSF grants No PHY-0240532, No PHY-0555522, and No CCF-0622100,
and by the NSF IGERT SELIM program at the University of Virginia.


\begin{thebibliography}{36}
\bibitem{note} Of course, from an experimental viewpoint, the phase
matching bandwidths will have to be considered. However, modern advances
in nonlinear photonic crystals make it quite feasible to implement
either case, as will be developed later.
\end{thebibliography}
\end{document}